\documentclass[%
 superscriptaddress,
 reprint,
 showpacs,preprintnumbers,
 nofootinbib,
 amsmath,amssymb,
 aps,
 prd,
 floatfix,
 longbibliography
]{revtex4-1}

\usepackage{graphicx}
\usepackage{xcolor}
\usepackage{graphicx}
\usepackage{subcaption} 
\usepackage{hyperref}
\hypersetup{
    colorlinks=true,
    linkcolor=blue,
    filecolor=blue,      
    urlcolor=blue,
    pdftitle={Neutrino-Nucleus Cross-Section Analysis with DUNE-PRISM}, 
    }
\makeatother

\usepackage{lipsum}
\usepackage{amssymb,amsmath,amstext,amsthm,amsfonts}
\usepackage[utf8]{inputenc}
\usepackage{float}
\usepackage{url}
\usepackage{soul}
\usepackage{esvect}
\usepackage{multirow}
\usepackage{enumitem}
\usepackage[export]{adjustbox}
\usepackage{braket}
\usepackage{comment}
\usepackage{caption}
\RequirePackage{xspace}
\usepackage{lineno}

\usepackage{orcidlink}

\usepackage{lineno}

\usepackage{todonotes}

\definecolor{LightCyan}{rgb}{0.88,1,1}
\usepackage{xcolor,colortbl}

\DeclareUnicodeCharacter{05D0}{\hebalef}
\DeclareUnicodeCharacter{05D1}{\hebbet}
\DeclareUnicodeCharacter{05D2}{\hebgimel}
\DeclareUnicodeCharacter{05D3}{\hebdalet}
\DeclareUnicodeCharacter{05D4}{\hebhe}
\DeclareUnicodeCharacter{05D5}{\hebvav}
\DeclareUnicodeCharacter{05D6}{\hebzayin}
\DeclareUnicodeCharacter{05D7}{\hebhet}
\DeclareUnicodeCharacter{05D8}{\hebtet}
\DeclareUnicodeCharacter{05D9}{\hebyod}
\DeclareUnicodeCharacter{05DA}{\hebfinalkaf}
\DeclareUnicodeCharacter{05DB}{\hebkaf}
\DeclareUnicodeCharacter{05DC}{\heblamed}
\DeclareUnicodeCharacter{05DD}{\hebfinalmem}
\DeclareUnicodeCharacter{05DE}{\hebmem}
\DeclareUnicodeCharacter{05DF}{\hebfinalnun}
\DeclareUnicodeCharacter{05E0}{\hebnun}
\DeclareUnicodeCharacter{05E1}{\hebsamekh}
\DeclareUnicodeCharacter{05E2}{\hebayin}
\DeclareUnicodeCharacter{05E3}{\hebfinalpe}
\DeclareUnicodeCharacter{05E4}{\hebpe}
\DeclareUnicodeCharacter{05E5}{\hebfinaltsadi}
\DeclareUnicodeCharacter{05E6}{\hebtsadi}
\DeclareUnicodeCharacter{05E7}{\hebqof}
\DeclareUnicodeCharacter{05E8}{\hebresh}
\DeclareUnicodeCharacter{05E9}{\hebshin}
\DeclareUnicodeCharacter{05EA}{\hebtav}

\newcommand{\FigureAutoref}[1]{\hyperref[#1]{Figure~\ref*{#1}}}

\begin{document}

\title{Towards mono-energetic virtual \(\nu\) beam cross-section measurements: A feasibility study of \(\nu\)-Ar interaction analysis with DUNE-PRISM}

%

%

\newcommand{\Albanysuny}{University of Albany, SUNY, Albany, NY 12222, USA}
\newcommand{\Almaty}{Institute of Nuclear Physics at Almaty, Almaty 050032, Kazakhstan
}
\newcommand{\Amsterdam}{University of Amsterdam, NL-1098 XG Amsterdam, The Netherlands}
\newcommand{\Antalya}{Antalya Bilim University, 07190 D\"o{\c s}emealtı/Antalya, Turkey}
\newcommand{\Antananarivo}{University of Antananarivo, Antananarivo 101, Madagascar}
\newcommand{\Antioquia}{University of Antioquia, Medell\'in, Colombia}
\newcommand{\AntonioNarino}{Universidad Antonio Nari\~no, Bogot\'a, Colombia}
\newcommand{\Argonne}{Argonne National Laboratory, Argonne, IL 60439, USA}
\newcommand{\Arizona}{University of Arizona, Tucson, AZ 85721, USA}
\newcommand{\Asuncion}{Universidad Nacional de Asunci\'on, San Lorenzo, Paraguay}
\newcommand{\Athens}{University of Athens, Zografou GR 157 84, Greece}
\newcommand{\Atlantico}{Universidad del Atl\'antico, Barranquilla, Atl\'antico, Colombia}
\newcommand{\Augustana}{Augustana University, Sioux Falls, SD 57197, USA}
\newcommand{\Bern}{University of Bern, CH-3012 Bern, Switzerland}
\newcommand{\Beykent}{Beykent University, Istanbul, Turkey}
\newcommand{\Birmingham}{University of Birmingham, Birmingham B15 2TT, United Kingdom}
\newcommand{\BolognaUniversity}{Universit\`a di Bologna, 40127 Bologna, Italy}
\newcommand{\Boston}{Boston University, Boston, MA 02215, USA}
\newcommand{\Bristol}{University of Bristol, Bristol BS8 1TL, United Kingdom}
\newcommand{\Brookhaven}{Brookhaven National Laboratory, Upton, NY 11973, USA}
\newcommand{\Bucharest}{University of Bucharest, Bucharest, Romania}
\newcommand{\CalBerkeley}{University of California Berkeley, Berkeley, CA 94720, USA}
\newcommand{\CalDavis}{University of California Davis, Davis, CA 95616, USA}
\newcommand{\CalIrvine}{University of California Irvine, Irvine, CA 92697, USA}
\newcommand{\CalLosangeles}{University of California Los Angeles, Los Angeles, CA 90095, USA}
\newcommand{\CalRiverside}{University of California Riverside, Riverside CA 92521, USA}
\newcommand{\CalSantabarbara}{University of California Santa Barbara, Santa Barbara, CA 93106, USA}
\newcommand{\Caltech}{California Institute of Technology, Pasadena, CA 91125, USA}
\newcommand{\Cambridge}{University of Cambridge, Cambridge CB3 0HE, United Kingdom}
\newcommand{\Campinas}{Universidade Estadual de Campinas, Campinas - SP, 13083-970, Brazil}
\newcommand{\CataniaUniversitadi}{Universit\`a di Catania, 2 - 95131 Catania, Italy}
\newcommand{\Catolica}{Universidad Cat\'olica del Norte, Antofagasta, Chile}
\newcommand{\CBPF}{Centro Brasileiro de Pesquisas F\'isicas, Rio de Janeiro, RJ 22290-180, Brazil}
\newcommand{\CEASaclay}{IRFU, CEA, Universit\'e Paris-Saclay, F-91191 Gif-sur-Yvette, France}
\newcommand{\CERN}{CERN, The European Organization for Nuclear Research, 1211 Meyrin, Switzerland}
\newcommand{\Charles}{Institute of Particle and Nuclear Physics of the Faculty of Mathematics and Physics of the Charles University, 180 00 Prague 8, Czech Republic }
\newcommand{\Chicago}{University of Chicago, Chicago, IL 60637, USA}
\newcommand{\ChungAng}{Chung-Ang University, Seoul 06974, South Korea}
\newcommand{\CIEMAT}{CIEMAT, Centro de Investigaciones Energ\'eticas, Medioambientales y Tecnol\'ogicas, E-28040 Madrid, Spain}
\newcommand{\Cincinnati}{University of Cincinnati, Cincinnati, OH 45221, USA}
\newcommand{\Cinvestav}{Centro de Investigaci\'on y de Estudios Avanzados del Instituto Polit\'ecnico Nacional (Cinvestav), Mexico City, Mexico}
\newcommand{\Colima}{Universidad de Colima, Colima, Mexico}
\newcommand{\ColoradoBoulder}{University of Colorado Boulder, Boulder, CO 80309, USA}
\newcommand{\ColoradoState}{Colorado State University, Fort Collins, CO 80523, USA}
\newcommand{\Columbia}{Columbia University, New York, NY 10027, USA}
\newcommand{\conida}{Comisi\'on Nacional de Investigaci\'on y Desarrollo Aeroespacial, Lima, Peru}
\newcommand{\Cti}{Centro de Tecnologia da Informacao Renato Archer, Amarais - Campinas, SP - CEP 13069-901}
\newcommand{\CUSB}{Central University of South Bihar, Gaya, 824236, India
}
\newcommand{\CzechAcademyofSciences}{Institute of Physics, Czech Academy of Sciences, 182 00 Prague 8, Czech Republic}
\newcommand{\CzechTechnical}{Czech Technical University, 115 19 Prague 1, Czech Republic}
\newcommand{\DannecyleVieux}{Laboratoire d'Annecy de Physique des Particules, Universit\'e Savoie Mont Blanc, CNRS, LAPP-IN2P3, 74000 Annecy, France}
\newcommand{\Daresbury}{Daresbury Laboratory, Cheshire WA4 4AD, United Kingdom}
\newcommand{\Dordt}{Dordt University, Sioux Center, IA 51250, USA}
\newcommand{\Drexel}{Drexel University, Philadelphia, PA 19104, USA}
\newcommand{\Duke}{Duke University, Durham, NC 27708, USA}
\newcommand{\Durham}{Durham University, Durham DH1 3LE, United Kingdom}
\newcommand{\Edinburgh}{University of Edinburgh, Edinburgh EH8 9YL, United Kingdom}
\newcommand{\EIA}{Universidad EIA, Envigado, Antioquia, Colombia}
\newcommand{\Eotvos}{E\"otv\"os Lor\'and University, 1053 Budapest, Hungary}
\newcommand{\erciyes}{Erciyes University, Kayseri, Turkey}
\newcommand{\FCULport}{Faculdade de Ci\^encias da Universidade de Lisboa - FCUL, 1749-016 Lisboa, Portugal}
\newcommand{\FederaldeAlfenas}{Universidade Federal de Alfenas, Po{\c c}os de Caldas - MG, 37715-400, Brazil}
\newcommand{\FederaldeGoias}{Universidade Federal de Goias, Goiania, GO 74690-900, Brazil}
\newcommand{\FederaldoABC}{Universidade Federal do ABC, Santo Andr\'e - SP, 09210-580, Brazil}
\newcommand{\FederaldoRio}{Universidade Federal do Rio de Janeiro, Rio de Janeiro - RJ, 21941-901, Brazil}
\newcommand{\Fermi}{Fermi National Accelerator Laboratory, Batavia, IL 60510, USA}
\newcommand{\Ferrarauniv}{University of Ferrara, Ferrara, Italy}
\newcommand{\Florida}{University of Florida, Gainesville, FL 32611-8440, USA}
\newcommand{\Floridastate}{Florida State University, Tallahassee, FL, 32306 USA}
\newcommand{\Fluminense}{Fluminense Federal University, 9 Icara\'i Niter\'oi - RJ, 24220-900, Brazil }
\newcommand{\Genova}{Universit\`a degli Studi di Genova, Genova, Italy}
\newcommand{\Georgian}{Georgian Technical University, Tbilisi, Georgia}
\newcommand{\Granada}{University of Granada \& CAFPE, 18002 Granada, Spain}
\newcommand{\GranSasso}{Gran Sasso Science Institute, L'Aquila, Italy}
\newcommand{\GranSassoLab}{Laboratori Nazionali del Gran Sasso, L'Aquila AQ, Italy}
\newcommand{\Grenoble}{University Grenoble Alpes, CNRS, Grenoble INP, LPSC-IN2P3, 38000 Grenoble, France}
\newcommand{\Guanajuato}{Universidad de Guanajuato, Guanajuato, C.P. 37000, Mexico}
\newcommand{\Harish}{Harish-Chandra Research Institute, Jhunsi, Allahabad 211 019, India}
\newcommand{\Hawaii}{University of Hawaii, Honolulu, HI 96822, USA}
\newcommand{\hkust}{Hong Kong University of Science and Technology, Kowloon, Hong Kong, China}
\newcommand{\Houston}{University of Houston, Houston, TX 77204, USA}
\newcommand{\Hyderabad}{University of  Hyderabad, Gachibowli, Hyderabad - 500 046, India}
\newcommand{\Idaho}{Idaho State University, Pocatello, ID 83209, USA}
\newcommand{\IFIC}{Instituto de F\'isica Corpuscular, CSIC and Universitat de Val\`encia, 46980 Paterna, Valencia, Spain}
\newcommand{\IGFAE}{Instituto Galego de F\'isica de Altas Enerx\'ias, University of Santiago de Compostela, Santiago de Compostela, 15782, Spain}
\newcommand{\ihep}{Institute of High Energy Physics, Chinese Academy of Sciences, Beijing, China}
\newcommand{\Iitk}{Indian Institute of Technology Kanpur, Uttar Pradesh 208016, India}
\newcommand{\Illinoisinstitute}{Illinois Institute of Technology, Chicago, IL 60616, USA}
\newcommand{\Imperial}{Imperial College of Science, Technology and Medicine, London SW7 2BZ, United Kingdom}
\newcommand{\IndGuwahati}{Indian Institute of Technology Guwahati, Guwahati, 781 039, India}
\newcommand{\IndHyderabad}{Indian Institute of Technology Hyderabad, Hyderabad, 502285, India}
\newcommand{\Indiana}{Indiana University, Bloomington, IN 47405, USA}
\newcommand{\INFNBologna}{Istituto Nazionale di Fisica Nucleare Sezione di Bologna, 40127 Bologna BO, Italy}
\newcommand{\INFNCatania}{Istituto Nazionale di Fisica Nucleare Sezione di Catania, I-95123 Catania, Italy}
\newcommand{\INFNFerrara}{Istituto Nazionale di Fisica Nucleare Sezione di Ferrara, I-44122 Ferrara, Italy}
\newcommand{\INFNFrascati}{Istituto Nazionale di Fisica Nucleare Laboratori Nazionali di Frascati, Frascati, Roma, Italy}
\newcommand{\INFNGenova}{Istituto Nazionale di Fisica Nucleare Sezione di Genova, 16146 Genova GE, Italy}
\newcommand{\INFNLecce}{Istituto Nazionale di Fisica Nucleare Sezione di Lecce, 73100 - Lecce, Italy}
\newcommand{\INFNMilanBicocca}{Istituto Nazionale di Fisica Nucleare Sezione di Milano Bicocca, 3 - I-20126 Milano, Italy}
\newcommand{\INFNMilano}{Istituto Nazionale di Fisica Nucleare Sezione di Milano, 20133 Milano, Italy}
\newcommand{\INFNNapoli}{Istituto Nazionale di Fisica Nucleare Sezione di Napoli, I-80126 Napoli, Italy}
\newcommand{\INFNPadova}{Istituto Nazionale di Fisica Nucleare Sezione di Padova, 35131 Padova, Italy}
\newcommand{\INFNPavia}{Istituto Nazionale di Fisica Nucleare Sezione di Pavia,  I-27100 Pavia, Italy}
\newcommand{\INFNPisa}{Istituto Nazionale di Fisica Nucleare Laboratori Nazionali di Pisa, Pisa PI, Italy}
\newcommand{\INFNRoma}{Istituto Nazionale di Fisica Nucleare Sezione di Roma, 00185 Roma RM, Italy}
\newcommand{\INFNRomavergata}{Istituto Nazionale di Fisica Nucleare Roma Tor Vergata , 00133 Roma RM, Italy}
\newcommand{\INFNSud}{Istituto Nazionale di Fisica Nucleare Laboratori Nazionali del Sud, 95123 Catania, Italy}
\newcommand{\Infntorino}{Istituto Nazionale di Fisica Nucleare, Sezione di Torino, Turin, Italy}
\newcommand{\Ingenieria}{Universidad Nacional de Ingenier\'ia, Lima 25, Per\'u}
\newcommand{\Insubria }{University of Insubria, Via Ravasi, 2, 21100 Varese VA, Italy}
\newcommand{\Iowa}{University of Iowa, Iowa City, IA 52242, USA}
\newcommand{\IowaState}{Iowa State University, Ames, Iowa 50011, USA}
\newcommand{\IPLyon}{Institut de Physique des 2 Infinis de Lyon, 69622 Villeurbanne, France}
\newcommand{\IPM}{Institute for Research in Fundamental Sciences, Tehran, Iran}
\newcommand{\IRLPPC}{Particle Physics and Cosmology International Research Laboratory	, Chicago IL,  60637 USA}
\newcommand{\ISTlisboa}{Instituto Superior T\'ecnico - IST, Universidade de Lisboa, 1049-001 Lisboa, Portugal}
\newcommand{\Ita}{Instituto Tecnol\'ogico de Aeron\'autica, Sao Jose dos Campos, Brazil}
\newcommand{\Iwate}{Iwate University, Morioka, Iwate 020-8551, Japan}
\newcommand{\Jacksonstate}{Jackson State University, Jackson, MS 39217, USA}
\newcommand{\Jawaharlal}{Jawaharlal Nehru University, New Delhi 110067, India}
\newcommand{\Jeonbuk}{Jeonbuk National University, Jeonrabuk-do 54896, South Korea}
\newcommand{\Jyvaskyla}{Jyv\"askyl\"a University, FI-40014 Jyv\"askyl\"a, Finland}
\newcommand{\Kansasstate}{Kansas State University, Manhattan, KS 66506, USA}
\newcommand{\Kavli}{Kavli Institute for the Physics and Mathematics of the Universe, Kashiwa, Chiba 277-8583, Japan}
\newcommand{\KEK}{High Energy Accelerator Research Organization (KEK), Ibaraki, 305-0801, Japan}
\newcommand{\KISTI}{Korea Institute of Science and Technology Information, Daejeon, 34141, South Korea}
\newcommand{\Kyiv}{Taras Shevchenko National University of Kyiv, 01601 Kyiv, Ukraine}
\newcommand{\Lancaster}{Lancaster University, Lancaster LA1 4YB, United Kingdom}
\newcommand{\LawrenceBerkeley}{Lawrence Berkeley National Laboratory, Berkeley, CA 94720, USA}
\newcommand{\LIP}{Laborat\'orio de Instrumenta{\c c}\~ao e F\'isica Experimental de Part\'iculas, 1649-003 Lisboa and 3004-516 Coimbra, Portugal}
\newcommand{\Liverpool}{University of Liverpool, L69 7ZE, Liverpool, United Kingdom}
\newcommand{\LosAlmos}{Los Alamos National Laboratory, Los Alamos, NM 87545, USA}
\newcommand{\Louisanastate}{Louisiana State University, Baton Rouge, LA 70803, USA}
\newcommand{\LpBordeaux}{Laboratoire de Physique des Deux Infinis Bordeaux - IN2P3, F-33175 Gradignan, Bordeaux, France, }
\newcommand{\Lucknow}{University of Lucknow, Uttar Pradesh 226007, India}
\newcommand{\Mainz}{Johannes Gutenberg-Universit\"at Mainz, 55122 Mainz, Germany}
\newcommand{\Manchester}{University of Manchester, Manchester M13 9PL, United Kingdom}
\newcommand{\Massinsttech}{Massachusetts Institute of Technology, Cambridge, MA 02139, USA}
\newcommand{\Medellin}{University of Medell\'in, Medell\'in, 050026 Colombia }
\newcommand{\Michigan}{University of Michigan, Ann Arbor, MI 48109, USA}
\newcommand{\Michiganstate}{Michigan State University, East Lansing, MI 48824, USA}
\newcommand{\MilanoBicocca}{Universit\`a di Milano Bicocca , 20126 Milano, Italy}
\newcommand{\MilanoUniv}{Universit\`a degli Studi di Milano, I-20133 Milano, Italy}
\newcommand{\Minnduluth}{University of Minnesota Duluth, Duluth, MN 55812, USA}
\newcommand{\Minntwin}{University of Minnesota Twin Cities, Minneapolis, MN 55455, USA}
\newcommand{\Mississippi}{University of Mississippi, University, MS 38677 USA}
\newcommand{\napoli}{Universit\`a degli Studi di Napoli Federico II , 80138 Napoli NA, Italy}
\newcommand{\Nikhef}{Nikhef National Institute of Subatomic Physics, 1098 XG Amsterdam, Netherlands}
\newcommand{\Niser}{National Institute of Science Education and Research (NISER), Odisha 752050, India}
\newcommand{\Northdakota}{University of North Dakota, Grand Forks, ND 58202-8357, USA}
\newcommand{\Northernillinois}{Northern Illinois University, DeKalb, IL 60115, USA}
\newcommand{\Northwestern}{Northwestern University, Evanston, Il 60208, USA}
\newcommand{\NotreDame}{University of Notre Dame, Notre Dame, IN 46556, USA}
\newcommand{\NoviSad}{University of Novi Sad, 21102 Novi Sad, Serbia}
\newcommand{\Ohiostate}{Ohio State University, Columbus, OH 43210, USA}
\newcommand{\OregonState}{Oregon State University, Corvallis, OR 97331, USA}
\newcommand{\Oxford}{University of Oxford, Oxford, OX1 3RH, United Kingdom}
\newcommand{\PacificNorthwest}{Pacific Northwest National Laboratory, Richland, WA 99352, USA}
\newcommand{\Padova}{Universt\`a degli Studi di Padova, I-35131 Padova, Italy}
\newcommand{\Panjab}{Panjab University, Chandigarh, 160014, India}
\newcommand{\Parissaclay}{Universit\'e Paris-Saclay, CNRS/IN2P3, IJCLab, 91405 Orsay, France}
\newcommand{\Parisuniversite}{Universit\'e Paris Cit\'e, CNRS, Astroparticule et Cosmologie, Paris, France}
\newcommand{\Parma}{University of Parma,  43121 Parma PR, Italy}
\newcommand{\Pavia}{Universit\`a degli Studi di Pavia, 27100 Pavia PV, Italy}
\newcommand{\Penn}{University of Pennsylvania, Philadelphia, PA 19104, USA}
\newcommand{\PennState}{Pennsylvania State University, University Park, PA 16802, USA}
\newcommand{\PhysicalResearchLaboratory}{Physical Research Laboratory, Ahmedabad 380 009, India}
\newcommand{\Pisa}{Universit\`a di Pisa, I-56127 Pisa, Italy}
\newcommand{\Pitt}{University of Pittsburgh, Pittsburgh, PA 15260, USA}
\newcommand{\Pontificia}{Pontificia Universidad Cat\'olica del Per\'u, Lima, Per\'u}
\newcommand{\PuertoRico}{University of Puerto Rico, Mayaguez 00681, Puerto Rico, USA}
\newcommand{\Punjab}{Punjab Agricultural University, Ludhiana 141004, India}
\newcommand{\QMUL}{Queen Mary University of London, London E1 4NS, United Kingdom
}
\newcommand{\Radboud}{Radboud University, NL-6525 AJ Nijmegen, Netherlands}
\newcommand{\Rice}{Rice University, Houston, TX 77005}
\newcommand{\Rochester}{University of Rochester, Rochester, NY 14627, USA}
\newcommand{\Royalholloway}{Royal Holloway College London, London, TW20 0EX, United Kingdom}
\newcommand{\Rutgers}{Rutgers University, Piscataway, NJ, 08854, USA}
\newcommand{\Rutherford}{STFC Rutherford Appleton Laboratory, Didcot OX11 0QX, United Kingdom}
\newcommand{\Salento}{Universit\`a del Salento, 73100 Lecce, Italy}
\newcommand{\santamarta}{Universidad del Magdalena, Santa Marta - Colombia}
\newcommand{\Sapienza}{Sapienza University of Rome, 00185 Roma RM, Italy}
\newcommand{\SergioArboleda}{Universidad Sergio Arboleda, 11022 Bogot\'a, Colombia}
\newcommand{\Sheffield}{University of Sheffield, Sheffield S3 7RH, United Kingdom}
\newcommand{\SLAC}{SLAC National Accelerator Laboratory, Menlo Park, CA 94025, USA}
\newcommand{\Southcarolina}{University of South Carolina, Columbia, SC 29208, USA}
\newcommand{\SouthDakotaSchool}{South Dakota School of Mines and Technology, Rapid City, SD 57701, USA}
\newcommand{\SouthDakotaState}{South Dakota State University, Brookings, SD 57007, USA}
\newcommand{\StonyBrook}{Stony Brook University, SUNY, Stony Brook, NY 11794, USA}
\newcommand{\SURF}{Sanford Underground Research Facility, Lead, SD, 57754, USA}
\newcommand{\Sussex}{University of Sussex, Brighton, BN1 9RH, United Kingdom}
\newcommand{\Syracuse}{Syracuse University, Syracuse, NY 13244, USA}
\newcommand{\Tecnologica }{Universidade Tecnol\'ogica Federal do Paran\'a, Curitiba, Brazil}
\newcommand{\TelAviv}{Tel Aviv University, Tel Aviv-Yafo, Israel}
\newcommand{\TexasAMcollege}{Texas A\&M University, College Station, Texas 77840}
\newcommand{\TexasAMcorpuscristi}{Texas A\&M University - Corpus Christi, Corpus Christi, TX 78412, USA}
\newcommand{\TexasArlington}{University of Texas at Arlington, Arlington, TX 76019, USA}
\newcommand{\Texasaustin}{University of Texas at Austin, Austin, TX 78712, USA}
\newcommand{\Toronto}{University of Toronto, Toronto, Ontario M5S 1A1, Canada}
\newcommand{\Tufts}{Tufts University, Medford, MA 02155, USA}
\newcommand{\Unifesp}{Universidade Federal de S\~ao Paulo, 09913-030, S\~ao Paulo, Brazil}
\newcommand{\UNIST}{Ulsan National Institute of Science and Technology, Ulsan 689-798, South Korea}
\newcommand{\UniversityCollegeLondon}{University College London, London, WC1E 6BT, United Kingdom}
\newcommand{\univkansas}{University of Kansas, Lawrence, KS 66045}
\newcommand{\UNMSM}{Universidad Nacional Mayor de San Marcos, Lima, Peru}
\newcommand{\ValleyCity}{Valley City State University, Valley City, ND 58072, USA}
\newcommand{\Vigo}{University of Vigo, E- 36310 Vigo Spain}
\newcommand{\VirginiaTech}{Virginia Tech, Blacksburg, VA 24060, USA}
\newcommand{\Warsaw}{University of Warsaw, 02-093 Warsaw, Poland}
\newcommand{\Warwick}{University of Warwick, Coventry CV4 7AL, United Kingdom}
\newcommand{\Wellesley}{Wellesley College, Wellesley, MA 02481, USA}
\newcommand{\Wichita}{Wichita State University, Wichita, KS 67260, USA}
\newcommand{\WilliamMary}{William and Mary, Williamsburg, VA 23187, USA}
\newcommand{\Wisconsin}{University of Wisconsin Madison, Madison, WI 53706, USA}
\newcommand{\Yale}{Yale University, New Haven, CT 06520, USA}
\newcommand{\Yerevan}{Yerevan Institute for Theoretical Physics and Modeling, Yerevan 0036, Armenia}
\newcommand{\York}{York University, Toronto M3J 1P3, Canada}
\affiliation{\Albanysuny}
\affiliation{\Almaty}
\affiliation{\Amsterdam}
\affiliation{\Antalya}
\affiliation{\Antananarivo}
\affiliation{\Antioquia}
\affiliation{\AntonioNarino}
\affiliation{\Argonne}
\affiliation{\Arizona}
\affiliation{\Asuncion}
\affiliation{\Athens}
\affiliation{\Atlantico}
\affiliation{\Augustana}
\affiliation{\Bern}
\affiliation{\Beykent}
\affiliation{\Birmingham}
\affiliation{\BolognaUniversity}
\affiliation{\Boston}
\affiliation{\Bristol}
\affiliation{\Brookhaven}
\affiliation{\Bucharest}
\affiliation{\CalBerkeley}
\affiliation{\CalDavis}
\affiliation{\CalIrvine}
\affiliation{\CalLosangeles}
\affiliation{\CalRiverside}
\affiliation{\CalSantabarbara}
\affiliation{\Caltech}
\affiliation{\Cambridge}
\affiliation{\Campinas}
\affiliation{\CataniaUniversitadi}
\affiliation{\Catolica}
\affiliation{\CBPF}
\affiliation{\CEASaclay}
\affiliation{\CERN}
\affiliation{\Charles}
\affiliation{\Chicago}
\affiliation{\ChungAng}
\affiliation{\CIEMAT}
\affiliation{\Cincinnati}
\affiliation{\Cinvestav}
\affiliation{\Colima}
\affiliation{\ColoradoBoulder}
\affiliation{\ColoradoState}
\affiliation{\Columbia}
\affiliation{\conida}
\affiliation{\Cti}
\affiliation{\CUSB}
\affiliation{\CzechAcademyofSciences}
\affiliation{\CzechTechnical}
\affiliation{\DannecyleVieux}
\affiliation{\Daresbury}
\affiliation{\Dordt}
\affiliation{\Drexel}
\affiliation{\Duke}
\affiliation{\Durham}
\affiliation{\Edinburgh}
\affiliation{\EIA}
\affiliation{\Eotvos}
\affiliation{\erciyes}
\affiliation{\FCULport}
\affiliation{\FederaldeAlfenas}
\affiliation{\FederaldeGoias}
\affiliation{\FederaldoABC}
\affiliation{\FederaldoRio}
\affiliation{\Fermi}
\affiliation{\Ferrarauniv}
\affiliation{\Florida}
\affiliation{\Floridastate}
\affiliation{\Fluminense}
\affiliation{\Genova}
\affiliation{\Georgian}
\affiliation{\Granada}
\affiliation{\GranSasso}
\affiliation{\GranSassoLab}
\affiliation{\Grenoble}
\affiliation{\Guanajuato}
\affiliation{\Harish}
\affiliation{\Hawaii}
\affiliation{\hkust}
\affiliation{\Houston}
\affiliation{\Hyderabad}
\affiliation{\Idaho}
\affiliation{\IFIC}
\affiliation{\IGFAE}
\affiliation{\ihep}
\affiliation{\Iitk}
\affiliation{\Illinoisinstitute}
\affiliation{\Imperial}
\affiliation{\IndGuwahati}
\affiliation{\IndHyderabad}
\affiliation{\Indiana}
\affiliation{\INFNBologna}
\affiliation{\INFNCatania}
\affiliation{\INFNFerrara}
\affiliation{\INFNFrascati}
\affiliation{\INFNGenova}
\affiliation{\INFNLecce}
\affiliation{\INFNMilanBicocca}
\affiliation{\INFNMilano}
\affiliation{\INFNNapoli}
\affiliation{\INFNPadova}
\affiliation{\INFNPavia}
\affiliation{\INFNPisa}
\affiliation{\INFNRoma}
\affiliation{\INFNRomavergata}
\affiliation{\INFNSud}
\affiliation{\Infntorino}
\affiliation{\Ingenieria}
\affiliation{\Insubria }
\affiliation{\Iowa}
\affiliation{\IowaState}
\affiliation{\IPLyon}
\affiliation{\IPM}
\affiliation{\IRLPPC}
\affiliation{\ISTlisboa}
\affiliation{\Ita}
\affiliation{\Iwate}
\affiliation{\Jacksonstate}
\affiliation{\Jawaharlal}
\affiliation{\Jeonbuk}
\affiliation{\Jyvaskyla}
\affiliation{\Kansasstate}
\affiliation{\Kavli}
\affiliation{\KEK}
\affiliation{\KISTI}
\affiliation{\Kyiv}
\affiliation{\Lancaster}
\affiliation{\LawrenceBerkeley}
\affiliation{\LIP}
\affiliation{\Liverpool}
\affiliation{\LosAlmos}
\affiliation{\Louisanastate}
\affiliation{\LpBordeaux}
\affiliation{\Lucknow}
\affiliation{\Mainz}
\affiliation{\Manchester}
\affiliation{\Massinsttech}
\affiliation{\Medellin}
\affiliation{\Michigan}
\affiliation{\Michiganstate}
\affiliation{\MilanoBicocca}
\affiliation{\MilanoUniv}
\affiliation{\Minnduluth}
\affiliation{\Minntwin}
\affiliation{\Mississippi}
\affiliation{\napoli}
\affiliation{\Nikhef}
\affiliation{\Niser}
\affiliation{\Northdakota}
\affiliation{\Northernillinois}
\affiliation{\Northwestern}
\affiliation{\NotreDame}
\affiliation{\NoviSad}
\affiliation{\Ohiostate}
\affiliation{\OregonState}
\affiliation{\Oxford}
\affiliation{\PacificNorthwest}
\affiliation{\Padova}
\affiliation{\Panjab}
\affiliation{\Parissaclay}
\affiliation{\Parisuniversite}
\affiliation{\Parma}
\affiliation{\Pavia}
\affiliation{\Penn}
\affiliation{\PennState}
\affiliation{\PhysicalResearchLaboratory}
\affiliation{\Pisa}
\affiliation{\Pitt}
\affiliation{\Pontificia}
\affiliation{\PuertoRico}
\affiliation{\Punjab}
\affiliation{\QMUL}
\affiliation{\Radboud}
\affiliation{\Rice}
\affiliation{\Rochester}
\affiliation{\Royalholloway}
\affiliation{\Rutgers}
\affiliation{\Rutherford}
\affiliation{\Salento}
\affiliation{\santamarta}
\affiliation{\Sapienza}
\affiliation{\SergioArboleda}
\affiliation{\Sheffield}
\affiliation{\SLAC}
\affiliation{\Southcarolina}
\affiliation{\SouthDakotaSchool}
\affiliation{\SouthDakotaState}
\affiliation{\StonyBrook}
\affiliation{\SURF}
\affiliation{\Sussex}
\affiliation{\Syracuse}
\affiliation{\Tecnologica }
\affiliation{\TelAviv}
\affiliation{\TexasAMcollege}
\affiliation{\TexasAMcorpuscristi}
\affiliation{\TexasArlington}
\affiliation{\Texasaustin}
\affiliation{\Toronto}
\affiliation{\Tufts}
\affiliation{\Unifesp}
\affiliation{\UNIST}
\affiliation{\UniversityCollegeLondon}
\affiliation{\univkansas}
\affiliation{\UNMSM}
\affiliation{\ValleyCity}
\affiliation{\Vigo}
\affiliation{\VirginiaTech}
\affiliation{\Warsaw}
\affiliation{\Warwick}
\affiliation{\Wellesley}
\affiliation{\Wichita}
\affiliation{\WilliamMary}
\affiliation{\Wisconsin}
\affiliation{\Yale}
\affiliation{\Yerevan}
\affiliation{\York}
\author{S.~Abbaslu} \affiliation{\IPM}
\author{A.~Abed Abud} \affiliation{\CERN}
\author{R.~Acciarri} \affiliation{\Fermi}
\author{L.~P.~Accorsi} \affiliation{\Tecnologica }
\author{M.~A.~Acero} \affiliation{\Atlantico}
\author{M.~R.~Adames} \affiliation{\Tecnologica }
\author{G.~Adamov} \affiliation{\Georgian}
\author{M.~Adamowski} \affiliation{\Fermi}
\author{C.~Adriano} \affiliation{\Campinas}
\author{F.~Akbar} \affiliation{\Rochester}
% ...
\collaboration{The DUNE Collaboration}
\noaffiliation

\begin{abstract}
\noindent Neutrino-nucleus cross-section measurements are critical for future neutrino oscillation analyses. However, our models to describe them require further refinement, and a deeper understanding of the underlying physics is essential for future neutrino oscillation experiments to realize their ambitious physics goals. Current neutrino cross-section measurements provide clear deficiencies in neutrino interaction modeling, but almost all are reported averaged over broad neutrino fluxes, rendering their interpretation challenging. 
Using the DUNE-PRISM concept (Deep Underground Neutrino Experiment Precision Reaction Independent Spectrum Measurement) — a movable near detector that samples multiple off-axis positions — neutrino interaction measurements can be used to construct narrow virtual fluxes (less than 100 MeV wide). These fluxes can be used to extract charged-current neutrino-nucleus cross sections as functions of outgoing lepton kinematics within specific neutrino energy ranges. 
Based on a dedicated simulation with realistic event statistics and flux-related systematic uncertainties, but assuming an almost-perfect detector, we run a feasibility study demonstrating how DUNE-PRISM data can be used to measure muon neutrino charged-current integrated and differential cross sections over narrow fluxes. We find that this approach enables a model independent reconstruction of powerful observables, including energy transfer, typically accessible only in electron scattering measurements, but that large exposures may be required for differential cross-section measurements with few-\% statistical uncertainties.  
\end{abstract}

\maketitle


\section{Introduction} \label{sec:introduction}

The next generation of accelerator-based neutrino oscillation experiments: DUNE~\cite{DUNE:2020ypp,DUNE:2020jqi} and Hyper-Kamiokande (Hyper-K)~\cite{Hyper-Kamiokande:2018ofw}, are poised to dramatically improve the characterization of neutrino oscillations. This will be achieved through measurements of neutrino interaction event rates in intense neutrino beams at near detectors located close to the beam production point (before oscillations occur) and again at far detectors, positioned many hundreds of kilometers away. 
The DUNE experiment, which aims to independently determine the leptonic CP-violating phase and the neutrino mass ordering without relying on input from other oscillation experiments, will use a broad neutrino flux with a peak neutrino energy of $\sim$3 GeV.

Both DUNE’s and Hyper-K’s abilities to achieve their target sensitivities will depend on our understanding of neutrino-nucleus interaction cross sections and the reconstruction of the incident neutrino energy from the measured final state particles~\cite{NuSTEC:2017hzk,Katori:2016yel}. Experimentally, the neutrino oscillation probability is determined from the measured neutrino interaction rate with atomic nuclei in detectors. This interaction rate arises from the product of the neutrino cross section, the energy resolution (smearing between true and reconstructed energy), and the incoming neutrino flux. Current models of neutrino interactions are constrained by experiment's near detectors but significant systematic uncertainties remain~\cite{NuSTEC:2017hzk, NOvA:2021nfi,T2K:2023mcm}. Even with very large exposure near detector data, neutrino interaction modeling uncertainties on the extrapolation of constraints from the near detector to the far detector can still play a very important role, and are likely to dominate future error budgets~\cite{NuSTEC:2017hzk}. Specialized measurements of neutrino-nucleus interaction cross sections from a wide variety of experiments demonstrate clear deficiencies in the modeling used for oscillation experiments (see e.g. Ref.\cite{Avanzini:2021qlx} for a review of this topic). However, these measurements are almost always reported as averaged over a broad neutrino energy flux with widths from a few 100 MeV to a few GeV. This is much wider than the width of features of interest in the cross section (for example the width of the quasi-elastic or resonant peaks that appear as a function of energy transfer or the hadronic invariant mass), rendering the interpretation of these measurements challenging. 

\begin{figure}
    \centering
    \includegraphics[width=0.7\linewidth]{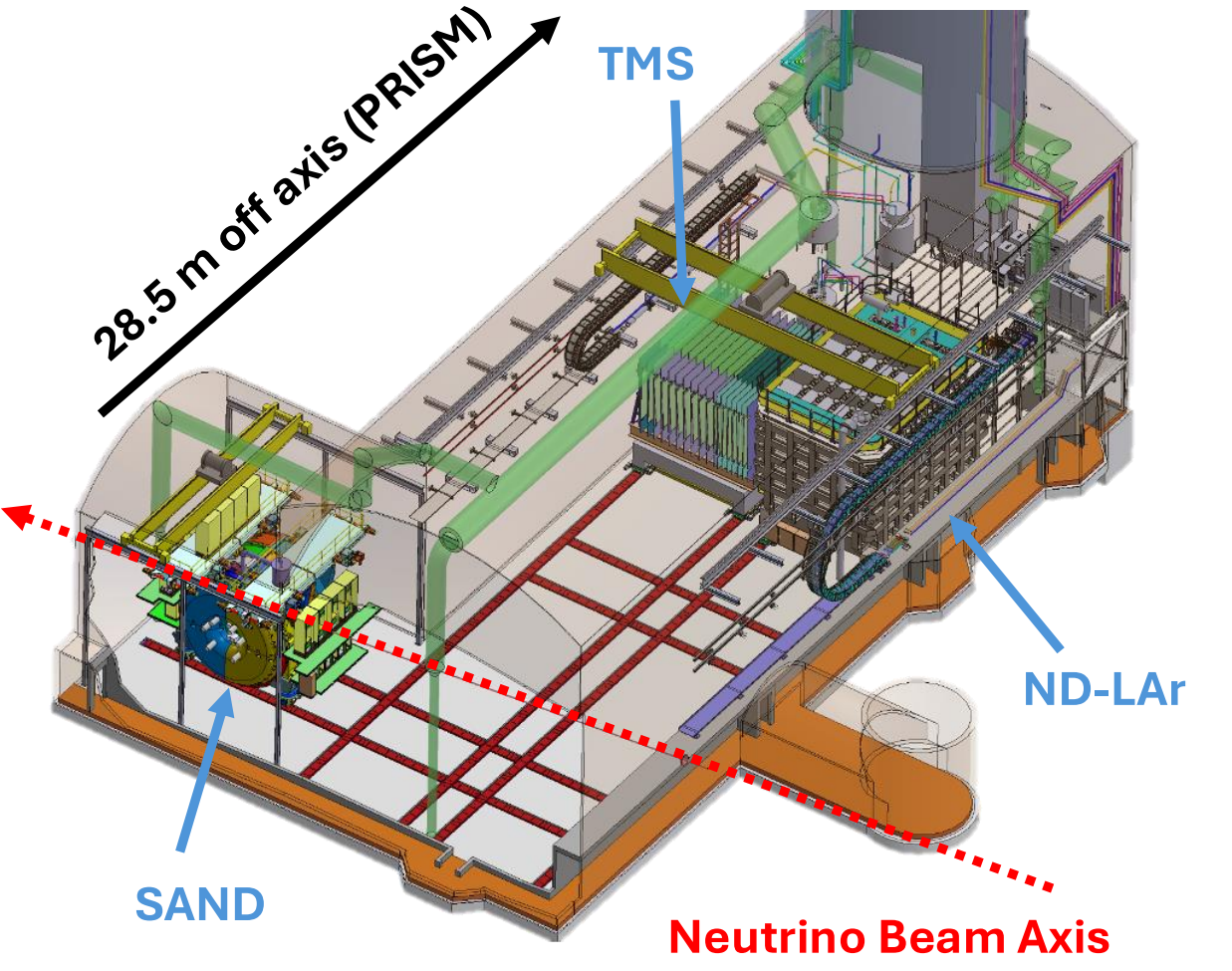}
    \caption{A schematic of the DUNE near detector. The SAND beam monitor is placed on-axis, whilst a Liquid Argon near detector (ND-LAr) and a muon spectrometer (TMS) are able to move from an on-axis position up to 28.5m off-axis. The beam axis and direction is indicated.
}
    \label{fig:PRISMdiagram}
\end{figure}

To constrain neutrino interactions and flux, DUNE employs a near detector, shown in \autoref{fig:PRISMdiagram}, consisting of three sub-detectors. Two of these sub-detectors utilize the PRISM concept, first suggested by the NuPRISM collaboration~\cite{nuPRISM:2014mzw}, allowing them to move with respect to the neutrino beam axis in a configuration called DUNE-PRISM~\cite{DUNE:2021tad}. At each position, or off-axis angle, a different flux distribution as a function of energy is sampled. 
Linear combinations of sampled fluxes at different off-axis positions allow for the creation of custom flux shapes. For instance, such a custom flux can be constructed to match the expected oscillated neutrino flux at the far detector.

Due to the absence of mono-energetic high-energy neutrino beams, a neutrino interaction model cannot typically be tested at individual neutrino energies. Instead, experiments tune the interaction models to reproduce their near-detector data. While informative, these flux integrated constraints are insufficient to ensure the models' accuracy for each value of energy or for the distribution of neutrino energies observed at the far detector. For precision measurements using a broad-energy neutrino beam, the extent to which near-detector data alone can constrain models is challenging to project, since the neutrino flux will differ significantly at the far detector due to oscillations. Thus, it is crucial to utilize as many neutrino interaction model constraints as possible using globally available data.

As part of ongoing efforts to address this challenge, various dedicated electron scattering experiments with targets and energies of interest to neutrino experiments are under operation and more are being designed~\cite{Ankowski:2022thw}. By knowing the incoming electron energy, such experiments allow a detailed and tailored exploration of the precise nuclear physics responsible for neutrino interaction uncertainties in analysis of neutrino oscillations. For example, electron scattering experiments have demonstrated the capability to test the energy reconstruction methods for various neutrino interaction final state topologies and compare their results to the actual beam energy and to electron interaction model predictions~\cite{Khachatryan:2021}. However, since electron-nuclei scattering interactions are dominated by the vector electromagnetic interaction, they cannot constrain the axial-vector component of the weak interaction. As a result, whilst measurements from electron scattering experiments provide essential constraints, a complete understanding of neutrino interactions requires further inputs from neutrino scattering experiments.

In this work, we investigate the feasibility of using the DUNE-PRISM concept to conduct neutrino measurements analogous to those performed in electron scattering by effectively constraining the energy of incoming neutrinos. 
By leveraging linear combinations of off-axis fluxes, we demonstrate a potential capability to generate narrow, custom flux spectra — referred to as virtual fluxes — and investigate how they can be used to measure neutrino interaction cross sections. Measurements made using this approach alleviate the major source of ambiguity in assessing the cause of current model-measurement disagreements, permitting robust measurements of observables usually reserved for electron scattering. This complements the few past measurements of mono-energetic neutrino scattering using pion or kaon decay-at-rest sources~\cite{JSNS2:2024bzf, MiniBooNE:2018dus}, which are limited to only very low neutrino energies (236 MeV for kaon decay at rest), far outside of the DUNE region of interest, and which have unknown flux normalisations (permitting shape-only cross-section measurements). The nuSCOPE collaboration propose to make similar measurements, but without using the PRISM concept, via a tagged neutrino beam~\cite{Acerbi:2025wzo}.

Within this work we consider muon-neutrino virtual-flux-averaged cross-section measurements of only lepton kinematics using a simple simulation in which we assume an almost perfect detector. We evaluate the broad expected statistical power of these  measurements, but we do not thoroughly optimise the virtual flux construction, which may allow improvements. The scope of this work is a simple first feasibility study to evaluate whether effective cross-section measurements over narrow fluxes can be made using DUNE-PRISM.

The simulation used for this analysis is described in \autoref{sec:simulation}, followed by the methods for constructing the virtual flux data sets in \autoref{sec:methods}, and the results in \autoref{sec:results}. These results are discussed in \autoref{sec:discuss} before concluding in \autoref{sec:concl}.

\section{Simulation}
\label{sec:simulation}

For this work, charged-current muon neutrino interactions on an argon nuclear target in the near detector are simulated using the NuWro 19.02.1 event generator~\cite{Golan:2012wx,Golan2012nuwro}. NuWro generates quasi-elastic interactions using the Benhar Spectral Function model~\cite{benhar1994spectral}, using an effective implementation for Argon~\cite{Ankowski:2005wi}, with optical potential corrections from Ref.~\cite{Ankowski:2014yfa} and a dipole axial form factor with the nucleon axial mass set to 1.03 GeV/$c^2$ (the default within NuWro). The 2p2h model is an implementation of the model from the Valencia group~\cite{gran2013neutrino,Nieves:2011pp}. Resonant pion production uses the Adler model~\cite{Adler:1975mt, ADLER1968189} considering only the dominant $\Delta$(1232) resonance. At higher momentum transfer, an inelastic model is used based on the GRV98 parton distribution functions~\cite{Gluck:1998xa} and PYTHIA6~\cite{Sjostrand:2006za}. NuWro is chosen due to both its event generation speed, as this work requires large simulations, and due to its use of a spectral function nuclear ground state, which is more realistic than the Fermi gas models prevalent in many other generators. In the context of this predominately qualitative feasibility study, it is not expected that the fine details of the neutrino interaction model will change the conclusions. 

One simulation is run using the flux for each of 58 different off-axis slices of 1 mrad, spanning angles between 0 and 3.32 degrees with respect to a position 574~m downstream of the detector along the central axis of the beamline's magnetic focusing horn (corresponding to the range of angles the near detector can span with the PRISM concept). More details on the flux modeling can be found in Refs.~\cite{DUNE:2021tad, DUNE:2020ypp, DUNE:2021mtg}. Note that in this work we only consider flux predictions using DUNE's nominal magnetic horn current (293 kA). At each off-axis position, 10 million interactions are generated. 

The event rates are weighted to match a run plan where each off-axis position is sampled for the same amount of time (assuming 1.1$\times$10$^{21}$ protons on target per year as in Ref.~\cite{DUNE:2021tad}), meaning events at large off-axis angles (where the flux is smaller, see \autoref{fig:FluxMatching}) are weighted down more. The resultant event rates are shown and discussed further in appendix \ref{app:evtrates}. Note that the run plan assumed in this work is different than DUNE's preliminary run plan, which allocates a little over half of the beam time on-axis ~\cite{DUNE:2021tad}. Therefore, one year of events used in the run plan considered in this study would be a subset of the events used in two years of the DUNE preliminary run plan. It should be noted that this chosen run plan is just an example modification to the preliminary plan that places more statistics in the off-axis angle samples but is not necessarily optimal. Future work may consider alternative run plans. 


As stated in \autoref{sec:introduction}, in this analysis we assume a perfect detector. This is to say that we do not account for detector smearing, efficiency effects or backgrounds arising from neutral current interactions, electron neutrino charged-current interactions, pileup or cosmic rays. However, since this study focuses solely on the outgoing muon kinematics (aside from a very loose cut on neutrino energy discussed later), these effects are not obviously large.

\section{Methods} \label{sec:methods}
\subsection{Constructing virtual flux data sets} \label{sec:virtualfluxes}
\subsubsection{Gaussian flux matching} \label{sec:matching}
\begin{figure*}[tb]
    \centering
    \includegraphics[width=0.9\textwidth]{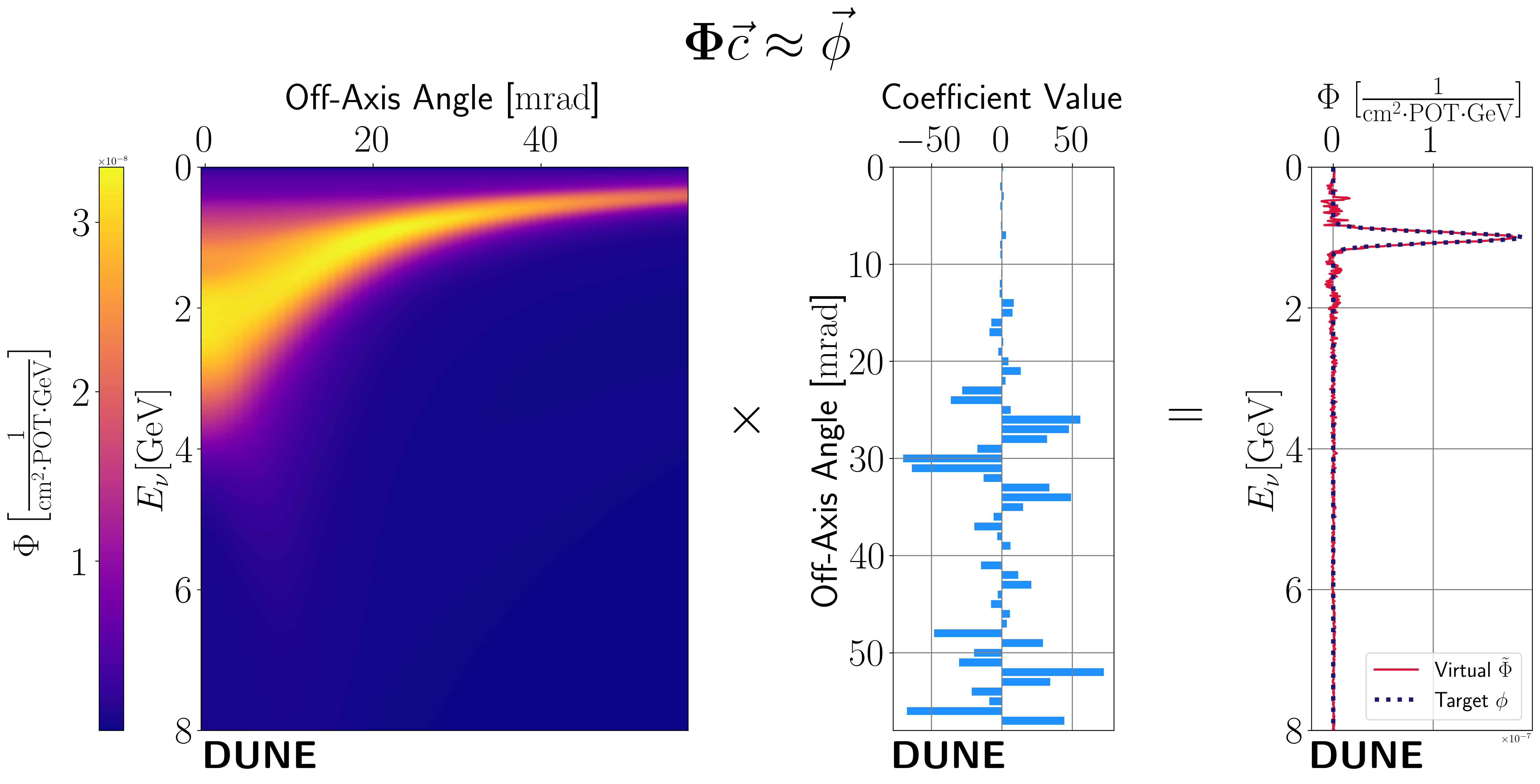}
    \caption{
        This visualization demonstrates the process of flux matching, as described in \autoref{sec:matching}. The flux matrix $\Phi$ is combined with a solution vector $\vec{c}$ to approximate the target flux $\phi(E_{\nu})$. (Left) A plot of the flux matrix \(\Phi\) used for this work, with 58 different bins of off-axis angle and 800 neutrino energy bins, each 10 MeV wide; 
        (Center) An ordinary least squares solution \(\vec{c}\) of \autoref{lincombomat} for a Gaussian target flux \(\phi(E_{\nu})\) centered at 1 GeV and with a standard deviation of 70 MeV; 
        (Right) A plot of both the target flux \(\phi(E_{\nu})\) and the virtual flux \(\tilde{\Phi}(E_{\nu})\) constructed by the solution in the middle plot.
    }
    \label{fig:FluxMatching}
\end{figure*}

As stated in \autoref{sec:introduction}, with DUNE-PRISM, measurements are performed at different detector positions relative to the neutrino beam. These measurements allow the construction of virtual fluxes \(\tilde{\Phi}(E_{\nu})\), which are linear combinations of the fluxes observed at different off-axis positions. We define \(\tilde{E}_{\nu}\) as the neutrino energy at which the virtual flux \(\tilde{\Phi}(E_{\nu})\) reaches its maximum.
A virtual flux can be tailored to approximate a desired flux shape, which we refer to as the target flux \(\phi(E_{\nu})\). Specifically, we construct virtual fluxes as follows:

\begin{equation} \label{lincomboflux}
\sum_{\alpha}^{}c_{\alpha}\Phi_{\alpha}(E_{\nu}) \approx \phi(E_{\nu}),
\end{equation}

\noindent where \(\Phi_{\alpha}(E_{\nu})\) is the flux at off-axis position \(\alpha\), and \(c_\alpha\) are the coefficients of the linear combination. For discrete energy bins indexed by \(i\), this system can be rewritten as:

\begin{equation} \label{lincombomat}
\Phi\vec{c} \approx \vec{\phi},
\end{equation}

\noindent where \(\Phi\) is a matrix with elements \(\Phi_{i\alpha}\), representing the flux at energy bin \(i\) and off-axis position \(\alpha\). The vector \(\vec{\phi}\) represents the target flux at energy bin \(i\), and \(\vec{c}\) is the coefficient vector that determines the contribution of each off-axis position. Note that we are free to choose the normalization of the target flux, since it cancels out when calculating cross sections and estimating uncertainties. \autoref{lincombomat} is visualized in \autoref{fig:FluxMatching}. Throughout this study we used a flux matrix with 800 energy bins from 0 to 8 GeV and 58 angle bins from 0 to 3.32 degrees.

Numerical methods can be used to find a \(\vec{c}\) which approximates \(\vec{\phi}\). Using the obtained coefficients, the virtual flux is defined as:

\begin{equation} \label{virtualflux}
\tilde{\Phi}(E_{\nu})=\sum_{\alpha} c_{\alpha}\Phi_{\alpha}(E_{\nu}),
\end{equation}

\noindent which serves as an approximation of \(\phi(E_{\nu})\). To generate a corresponding data set, events from different off-axis positions are reweighted according to \(c_{\alpha}\), yielding a virtual event rate:

\begin{equation} \label{virtualeventrates}
\tilde{N}=\sum_{\alpha}c_{\alpha}N_{\alpha},
\end{equation}

\noindent where \(N_{\alpha}\) represents the observed event rate at off-axis position \(\alpha\). Note that while \(N_{\alpha}\) are measured numbers of events, and therefore are natural numbers, \(\tilde{N}\) is a real number, calculated by reweighting the observed numbers of events \(N_{\alpha}\) by real weights \(c_{\alpha}\), and so \(\tilde{N}\) can even be negative.

The statistical uncertainty on the virtual event rate \(\tilde{N}\) is
\begin{equation} \label{uncertainty}
\frac{\Delta \tilde{N}}{\tilde{N}} = \frac{\sqrt{\sum_{\alpha} c_{\alpha}^2 N_{\alpha}}}{\left| \sum_{\alpha} c_{\alpha} N_{\alpha} \right|},
\end{equation}
and it is dependent on both the coefficients and the event rates at each off-axis angle.

Solving \autoref{lincombomat} using an ordinary least squares approach often results in high statistical uncertainties. Large positive and negative coefficients may cancel when computing the expected event rate but still contribute additively to the variance, as can be deduced from \autoref{uncertainty}. To mitigate this issue, we apply Tikhonov regularization~\cite{Tikhonov1995}, adding a penalty term proportional to the total \(L_2\) norm \(||\vec{c}||^2\). This suppresses large overall coefficient values, reducing statistical fluctuations at the cost of slightly broader virtual fluxes. 
Although regularization reduces the magnitude of the coefficients \(c_{\alpha}\), its main effect is a reduction in the numerator of the right hand side of \autoref{uncertainty}. The denominator remains approximately fixed due to the flux-matching constraint, while the numerator decreases as regularization pulls positive and negative \(c_{\alpha}\) toward smaller absolute values. These still cancel in the sum but contribute less to the variance, leading to a monotonic decrease in statistical uncertainty with increasing regularization strength. 

\begin{figure*} [tb]
  \centering
  \begin{subfigure}{0.49\textwidth}
    \centering
    \includegraphics[width=\linewidth]{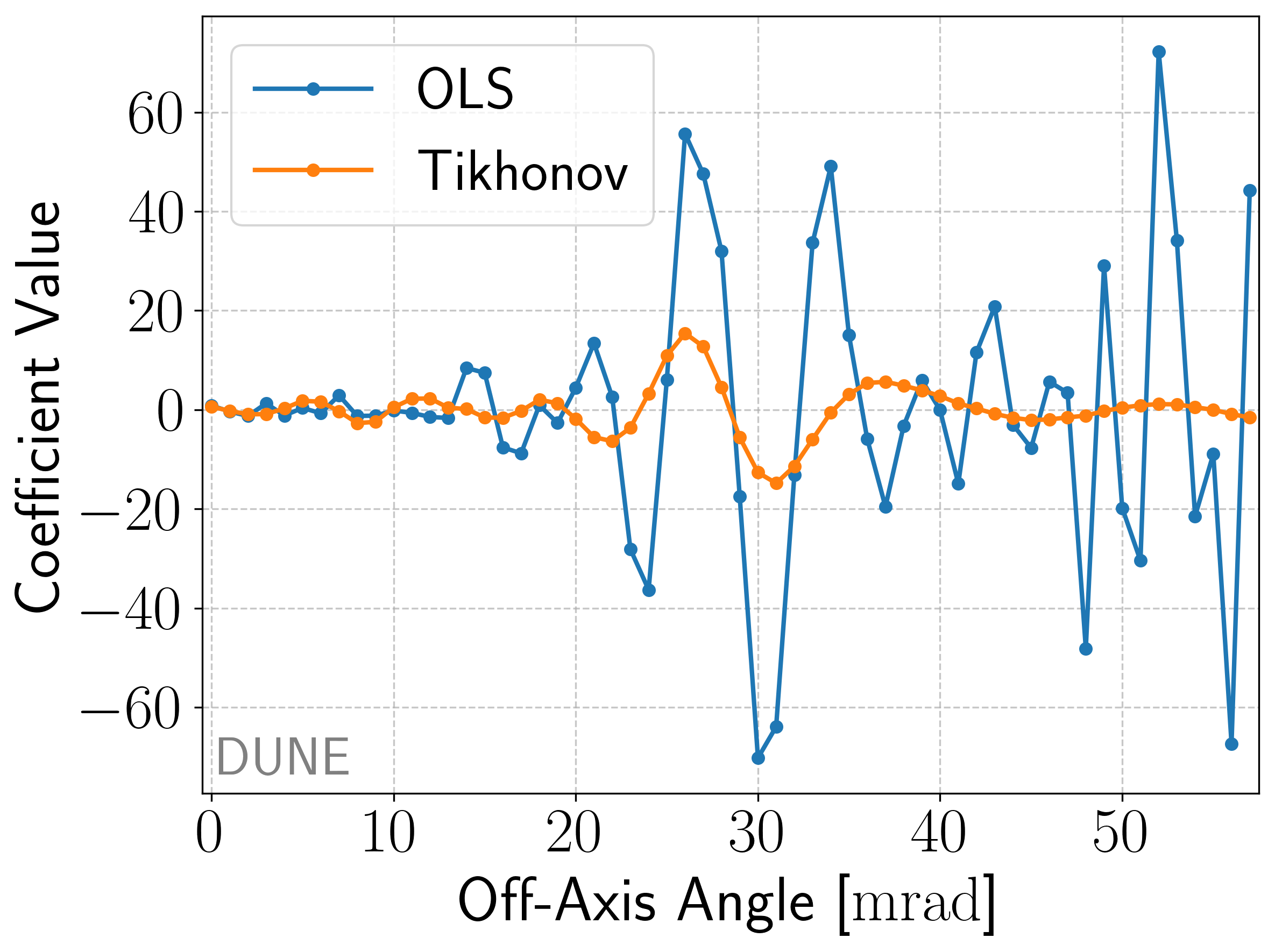}
    \caption{}
    \label{fig:RegExamplecoeffs}
  \end{subfigure}
  \hfill
  \begin{subfigure}{0.49\textwidth}
    \centering
    \includegraphics[width=\linewidth]{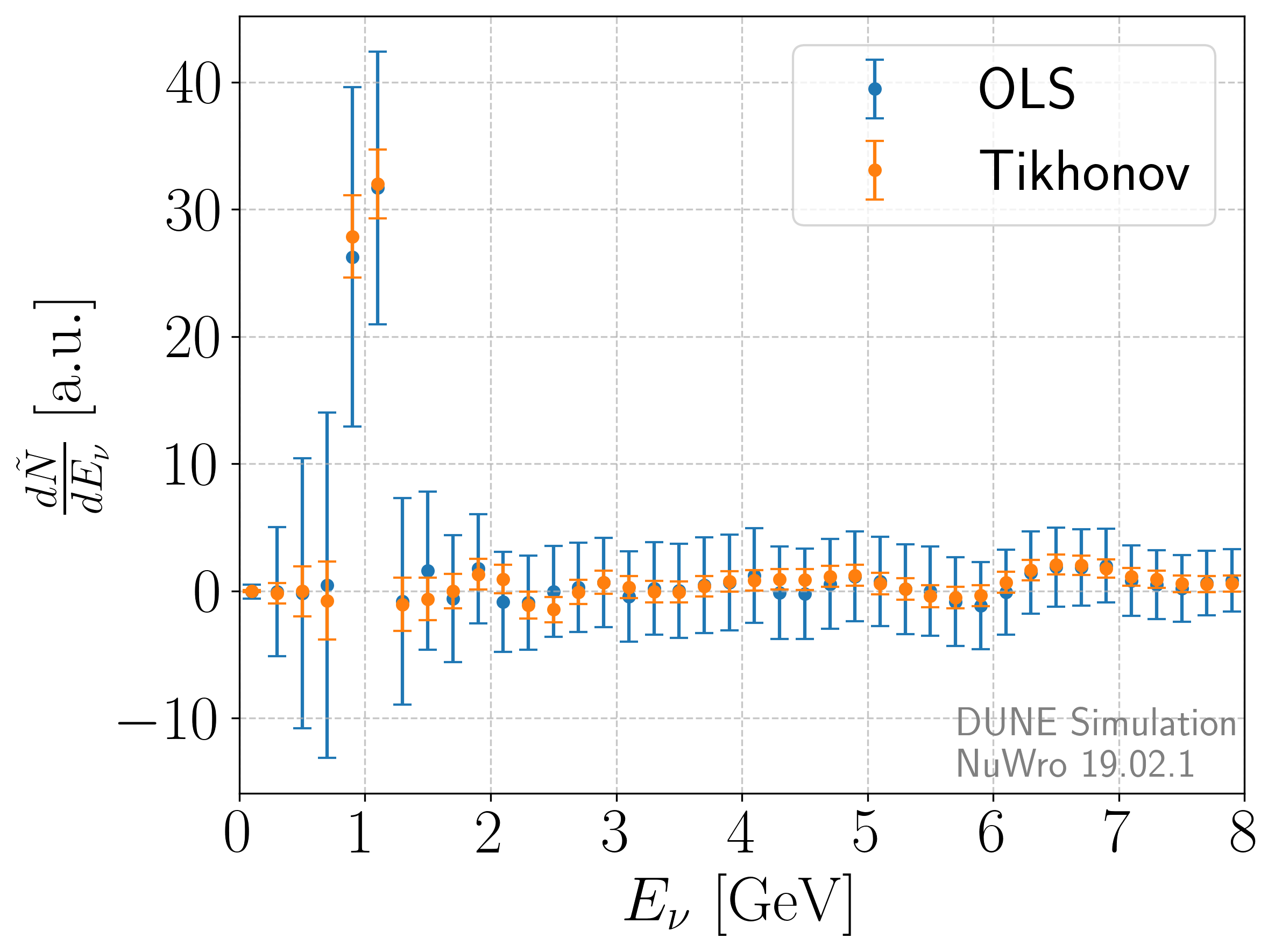}
    \caption{}
    \label{fig:RegExampleEventRate}
  \end{subfigure}
    \caption{
    (a) Coefficients as a function of the off-axis position index for the ordinary least squares (OLS) solution and the Tikhonov-regularized solution, corresponding to the virtual flux construction based on a Gaussian target flux with $\mu=1$ GeV and $\sigma=70$ MeV, as defined in the linear combination described in \autoref{lincombomat}. 
    (b) Virtual event rates as a function of true neutrino energy using virtual fluxes based on the same Gaussian target flux. The event rates constructed based on an ordinary least squares solution of \autoref{lincombomat} have $\sigma=71$ MeV (estimated from a Gaussian fit to the virtual event rate), and the event rates constructed based on a Tikhonov-regularized solution have $\sigma=78$ MeV.
    }
  \label{fig:Regularisation}
\end{figure*}

In general, regularization parameters can be determined using methods such as the L-curve method \cite{LCurve}. However, in this analysis, they were chosen manually, based on the known properties of the off-axis flux distribution and the target flux. An illustration of the application of regularization is shown in \autoref{fig:Regularisation}, demonstrating that, while the obtained virtual flux is wider than the target flux, the statistical uncertainties associated with the corresponding event rates are significantly reduced.

\subsubsection{Uncertainties}
Uncertainties are considered by producing two separate sets of ensembles (``toys'') of event rate histograms, based on statistical and systematic uncertainties. Statistical uncertainty toys were randomly drawn from Poisson distributions for each event rate bin.

To produce flux systematic uncertainty toys, we consider the 13 beamline systematic uncertainty components, which correspond to variations in the target density, water layer thickness, magnetic focusing horn current and position (4 position components in total), decay pipe radius, and proton beam position, angle (2 angular components in total), and width, as used in Ref.~\cite{DUNE:2021tad}, as well as proton-on-target (POT) counting. 
Uncertainties in the flux due to unknown cross sections for production of the parent hadrons of neutrinos are also considered, following the same approach described in Ref.~\cite{DUNE:2020jqi}. Variations of the incoming neutrino flux at each off-axis position are considered under variations of hadron-production cross sections to build a covariance matrix correlating their impact across bins of neutrino energy and off-axis angle. The vast majority of the variance encoded in the matrix can be extracted through a relatively small number of independent principal components (PCs). This analysis includes the 20 most significant independent hadron production components, as adding more was found to have a negligible impact on the uncertainty.  

To generate the \(n\)-th toy data set, events in off-axis angle bin \(\alpha\) and energy bin \(i\) were reweighted according to:
\begin{equation}
    w_{n\alpha i}=1+\sum^{\# \text{ PCs}}_{j=1}{X_{nj}\sigma_{j\alpha i}},
\end{equation}
where $X_{nj}$ is sampled from a Gaussian distribution with a mean of zero and a standard deviation of one and $\sigma_{j\alpha i}$ is the relative variation of an event in off-axis angle bin \(\alpha\) and neutrino energy bin \(i\), when varying just the \(j\)-th uncertainty principal component (PC).

Note that the input flux uncertainty estimates were coarsely binned relative to our simulated data, both in energy and off-axis angle. This required us to redistribute generated events into the binning used for the flux uncertainty.
Note that the input flux uncertainty estimates were coarsely binned relative to our simulated data, both in energy and off-axis angle. Specifically, the input flux covariance matrix spans 34 off-axis positions, each with between 14 and 32 discrete energy bins, whereas our simulated events are distributed across 58 off-axis angles with practically continuous energy values (up to NuWro's numerical resolution). This mismatch required us to redistribute the generated events into the coarser binning used for the flux uncertainty evaluation.
Since measurements with different virtual fluxes use the same sets of data (although weighted differently), it is important to consider correlations between different virtual flux measurements induced by both systematic and statistical uncertainties. By using the same set of toys for different virtual fluxes such correlations can be evaluated. For each of our main results in \autoref{sec:results} we present statistical, systematic and total uncertainty correlations (for example, \autoref{fig:intxseccorr}).

Due to the large normalization uncertainty — shifting all bins coherently — and small shape uncertainty implicit in flux uncertainties (for example, see Ref.~\cite{Koch:2020oyf}), it is unsurprising that simulations show strong bin-to-bin correlations, which lead to significant normalization uncertainties in event rate distributions. To calculate the uncertainties regarding the distribution's shape and normalization uncertainties separately, we used the Norm-Shape, ``NS'' decomposition scheme from Ref.~\cite{Chakrani:2023htw}. We normalize each distribution in order to calculate the shape-only uncertainty, and keep track of the normalization uncertainty using the last bin. For each toy, the full distribution of bins across all the plotted variable's bins 
and virtual flux centers \(\tilde{E_{\nu}}\), denoted by \(N_{i}\), with \(n\) total bins, is transformed with the following bijective mapping:

\begin{equation}
f_i=\left\{\begin{matrix}
\frac{N_i}{\sum_k N_k}, & 1\leq i\leq n-1\\ \sum_k N_k & i=n.
\end{matrix}\right. 
\end{equation}

The covariance is then calculated for each pair of bins, so the relative shape uncertainty is the standard deviation of \(f(N_i\)) across toys for \(i<n\). The relative normalization uncertainty is the standard deviation of \(f(N_n)\).
The systematic uncertainty is decomposed into normalization and shape components to enable separate inspection of their contributions. However, these components are correlated, and therefore their quadrature sum does not recover the full systematic uncertainty. To provide a more complete representation, both components are shown individually, alongside the total uncertainty obtained by combining the full systematic uncertainty (including correlations) with the statistical uncertainty in quadrature.


\subsection{Cross-section measurements with virtual flux data sets}
\label{sec:xsecwithvflux}

With a virtual flux defined by the set of coefficients \(\vec{c}\) and our event simulation,  virtual measurements can be obtained. This is done by adding each off-axis measurement histogram along with its corresponding coefficient as a weight. We simulated a measurement of the integrated cross section \(\sigma(E_{\nu})\) by building six narrow Gaussian virtual fluxes \(\tilde{\Phi}(E_{\nu})\) each centered around a \(\tilde{E}_{\nu}\) (the mean energy of the target flux) between 0.5~GeV and 1.75~GeV in steps of 0.25~GeV. The standard deviation of the target fluxes is set to 70 MeV (which is comparable with the expected width of features in the cross sections we aim to extract). We limit the analysis to 1.75~GeV, as it becomes increasingly challenging to build smooth, thin virtual fluxes centered at higher energies without applying very large coefficients (which would propagate to very large statistical uncertainties) due to limited coverage of PRISM off-axis spectrum.  While this threshold was chosen conservatively, it could be further optimized depending on the desired balance between flux fidelity and energy reach. 

It is important to note that this study does not rely on the virtual fluxes being an exact representation of the target fluxes. The goal is to produce effective fluxes that are thin and well known. The utility of cross-section measurements is not affected by the virtual fluxes being shifted, skewed or having a slightly different standard deviation with respect to the target flux. It is possible that a Gaussian choice of target flux is sub-optimal, and so future work could consider other target flux shapes (or explore analysis methods in which a specific target flux is not necessary).

Cross sections averaged over a virtual flux can be calculated through the following relation:

\begin{equation} \label{virtualxsec}
\langle \tilde{\sigma} \rangle=\frac{\tilde{N}}{\mathcal{E}\cdot\varepsilon\cdot N_{\rm targets}\int{\tilde{\Phi}(E_{\nu})}dE_{\nu}}, 
\end{equation}

\noindent where $\tilde{N}$ is the expected event rate in a virtual flux as defined in \autoref{virtualeventrates}. $\mathcal{E}$ represents the beam exposure (i.e. the number of protons on target considered) and $\varepsilon$ is the selection efficiency (always one for this analysis). A derivation of this result is provided in Appendix~\ref{app:Virtual flux equation}. The number of target nuclei, \(N_{\rm targets}\), is proportional to the mass of the DUNE-PRISM detector's fiducial volume. We assume a fiducial volume of \(2 \times 3 \times 0.574~\mathrm{m}^3\), corresponding to a total mass of 4.8~tonnes of liquid argon. Here it is important to note that we apply the coefficients to the event rate (i.e. by using $\tilde{N}$) rather than using a linear combination of separate cross-section measurements, as is discussed in appendix~\ref{app:Virtual flux equation}. Note that the virtual flux scaling is determined by the calculated coefficients (see \autoref{virtualflux}). The normalization of the virtual flux is arbitrary and carries no physical meaning, but it must be included consistently so that it cancels properly when converting event rates into cross section measurements. In this work we consider measurements of the muon neutrino charged-current inclusive cross section.

\subsubsection{Integrated cross-section measurements} \label{sec:intxsecmethods}


A challenge in extracting a virtual-flux-integrated cross section arises from small, localized ``bumps'' that often appear in the tails of the virtual fluxes. These are artifacts of the limited span of the flux matrix: since it is constructed from a finite set of off-axis input fluxes, it cannot exactly reproduce all possible smooth energy spectra. These bumps are especially problematic because the neutrino interaction cross section $\sigma(E_{\nu})$ generally increases with energy in the relevant range, so even a small excess in the high-energy tail of the flux can be amplified into a disproportionately large contribution to the event rate, as seen near $\sim$6.5\,GeV in \autoref{fig:RegExampleEventRate}. Consequently, integrating over such a flux can yield a significantly biased cross-section estimate — either positively or negatively — relative to what would be obtained using an idealized, smooth target flux such as a Gaussian.


To suppress contributions from these unwanted features of virtual fluxes, we opt to apply a loose cut of 4~GeV on the sum of visible outgoing particle energies (i.e. using the sum of $E_{\mu}$, the outgoing muon energy and $E_{avail}$, the sum of proton and charged pion kinetic energy, plus neutral pion, and photon total energy, introduced in Ref.~\cite{MINERvA:2015ydy} and further explored in Ref.~\cite{Wilkinson:2022dyx}). The sum $E_{\mu}+E_{avail}$ serves as a proxy for neutrino energy with relatively sufficiently small smearing.
The migration of events with true neutrino energy larger than 4~GeV across the 4~GeV cut is small when the cut lies well above the peak of the virtual flux; in our simulation, the subsequent background-to-signal ratio remains between 1.6\% and 5.0\% for all fluxes except one, where a prominent bump near 4~GeV inflates the ratio to 18.0\%. However, this has no effect on our results, as the background subtraction in our simulation is exact. For such cases, in a more sophisticated analysis, the threshold can easily be adjusted to a different value (e.g. 3 GeV) while still remaining sufficiently separated from the bulk of the virtual flux.

Such a cut must be accompanied by a redefinition of the cross section we measure to be flux averaged only over the range of neutrino energy up to the cut value. In addition, a small background subtraction is required for events that pass the cut but in fact have true neutrino energies above 4 GeV. This can be written as:

\begin{equation} \label{virtualxsec_cut}
\langle \tilde{\sigma}_{\rm cut} \rangle=\frac{\tilde{N}-\tilde{B}}{\mathcal{E}\cdot\varepsilon\cdot N_{\rm targets}\int^{E'}_{0}{\tilde{\Phi}(E_{\nu})}dE_{\nu}}, 
\end{equation}

\noindent where \(\tilde{B}=\sum_{\alpha}c_{\alpha}B_{\alpha}\) and \(B_{\alpha}\) is the expected background rate at an off-axis position \(\alpha\) for a threshold set to \(E'\) (4 GeV in our case), and \(\varepsilon\) should incorporate the simulated efficiency of the cut. Note that as reconstruction-induced smearing is excluded, the efficiency of the cut is still 100\%, as kinematically the neutrino energy proxy is always less than the true neutrino energy.  Whilst the background subtraction introduces a model dependence, the looseness of our cut ensures that the background introduced by smearing over the cut boundary was found to be on the scale of no more than a few percent. As discussed, the exact threshold can be optimized to minimize the amount of background, depending on the particular features of a virtual flux, based on the expected background and at the cost of a slightly increased interaction model dependence. We note that since the background subtraction is taken directly from the simulation and we do not consider any uncertainties in neutrino interaction modeling, this procedure is equivalent to simply cutting on true neutrino energy.


\subsubsection{Differential cross-section measurements}


We additionally extract the differential cross section as a function of the reconstructed energy transfer \(\frac{d\sigma}{d\omega_{\mathrm{reco}}}\) where:

\begin{equation}
\label{eq_omegareco}
    \omega_{\mathrm{reco}}=\tilde{E}_{\nu} - E_{\mathrm{lepton}},
\end{equation}

\noindent where $\tilde{E}_{\nu}$ is the aforementioned mean energy of the target flux and $E_{lepton}$ is the energy of the outgoing lepton. In the limit of a narrow virtual flux, \(\tilde{E}_{\nu}\approx E_{\nu}\), such that $\omega_{\mathrm{reco}} \approx \omega_{\mathrm{true}}$, with the true energy transfer defined as 

\begin{equation}
\label{eq_omegatrue}
    \omega_{\mathrm{true}}=E_{\nu} - E_{\mathrm{lepton}}.
\end{equation}

Energy transfer is of particular interest because different reaction channels and key nuclear effects, which are relevant for DUNE oscillation analyses, are strongly energy transfer dependent~\cite{NuSTEC:2017hzk, Ankowski:2022thw, Ankowski:2020qbe}. It is often measured as a key observable for nuclear physics analyses in electron scattering experiments but usually remains inaccessible to neutrino experiments due to the broadband nature of neutrino beams rendering $\omega_{\mathrm{reco}}$ very smeared with respect to $\omega_{\mathrm{true}}$. However, it should be noted that electron scattering measurements usually measure energy transfers for an approximately fixed scattering angle, resulting in a distribution of energy transfer with distinct peaks for different interaction channels. This corresponds to a narrow scan through the 2D space of energy and three momentum transfer, which reveals a multitude of features of inclusive cross sections. Unfortunately, significant  statistical uncertainties resulting from the narrow Gaussian flux matching makes tight angular restrictions impractical in the current analysis. While energy transfer remains an interesting observable and serves as a proof of principle of the methods introduced, appendix~\ref{app:Hadronic} also considers the hadronic invariant mass, which can help disentangle different interaction channels while reducing reliance on a fixed scattering angle.  

To produce this measurement, for each of the aforementioned six virtual fluxes, a histogram of event rates as a function of \(\omega_{\mathrm{reco}}\) is generated. The event rate is then transformed to a virtual flux-averaged differential cross-section measurement using \autoref{virtualxsec}, with an additional term to account for the bin width in \(\omega_{\mathrm{reco}}\). 




Another advantage of \(\omega_{\mathrm{reco}}\) is in an opportunity to use a minimally model dependent unfolding procedure to \(\omega_{\mathrm{true}}=E_{\nu}-E_{\text{lepton}}\), which is discussed in appendix~\ref{app:Unfolding}.

\section{Results}

\label{sec:results}
\subsection{Integrated cross-section measurements}
\begin{figure*} [tb]
  \centering
  \begin{subfigure}{0.49\textwidth}
    \centering
    \includegraphics[width=\linewidth]{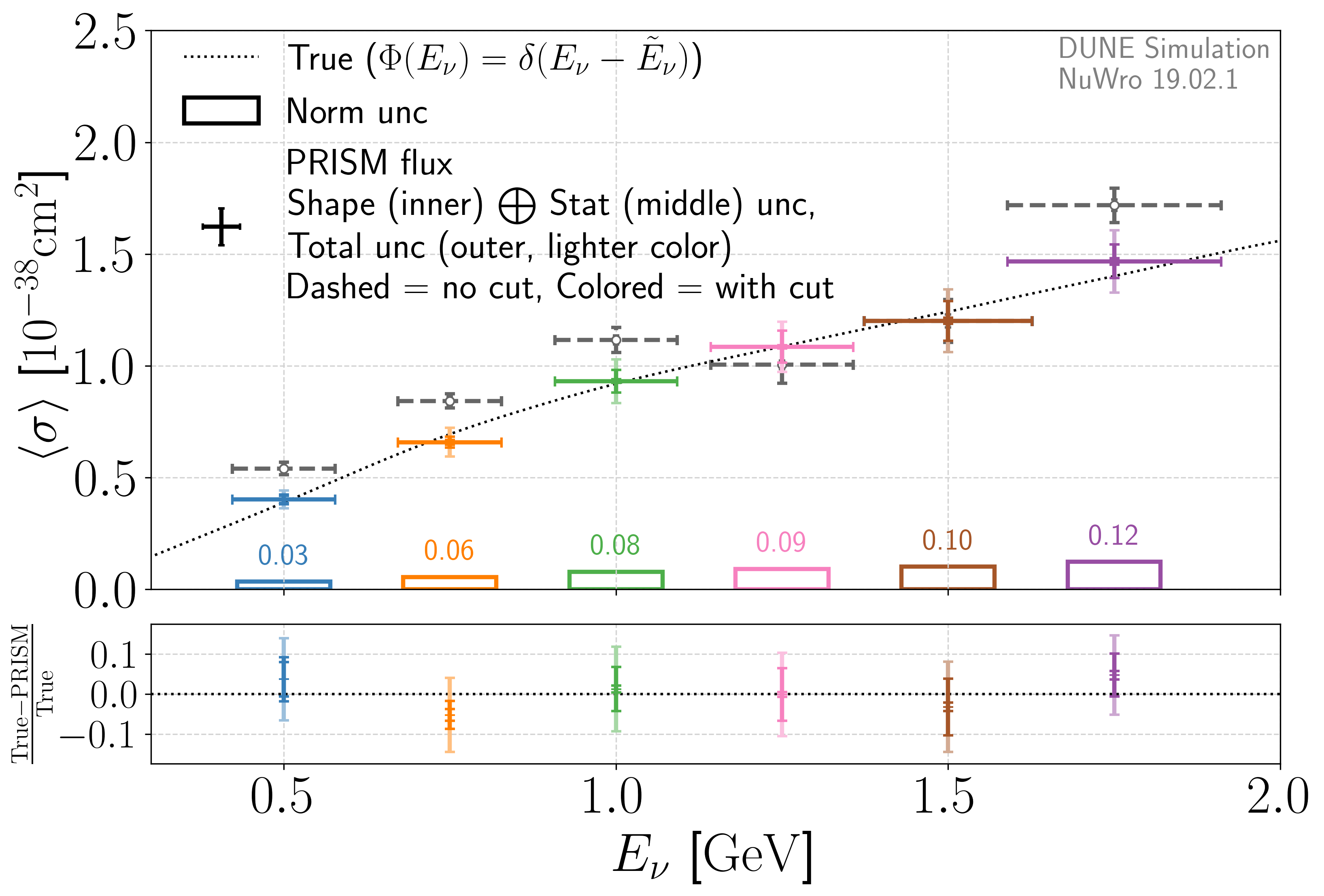}
    \caption{}
    \label{fig:IntXSecSubfig1}
  \end{subfigure}
  \hfill
  \begin{subfigure}{0.49\textwidth}
    \centering
    \includegraphics[width=\linewidth]{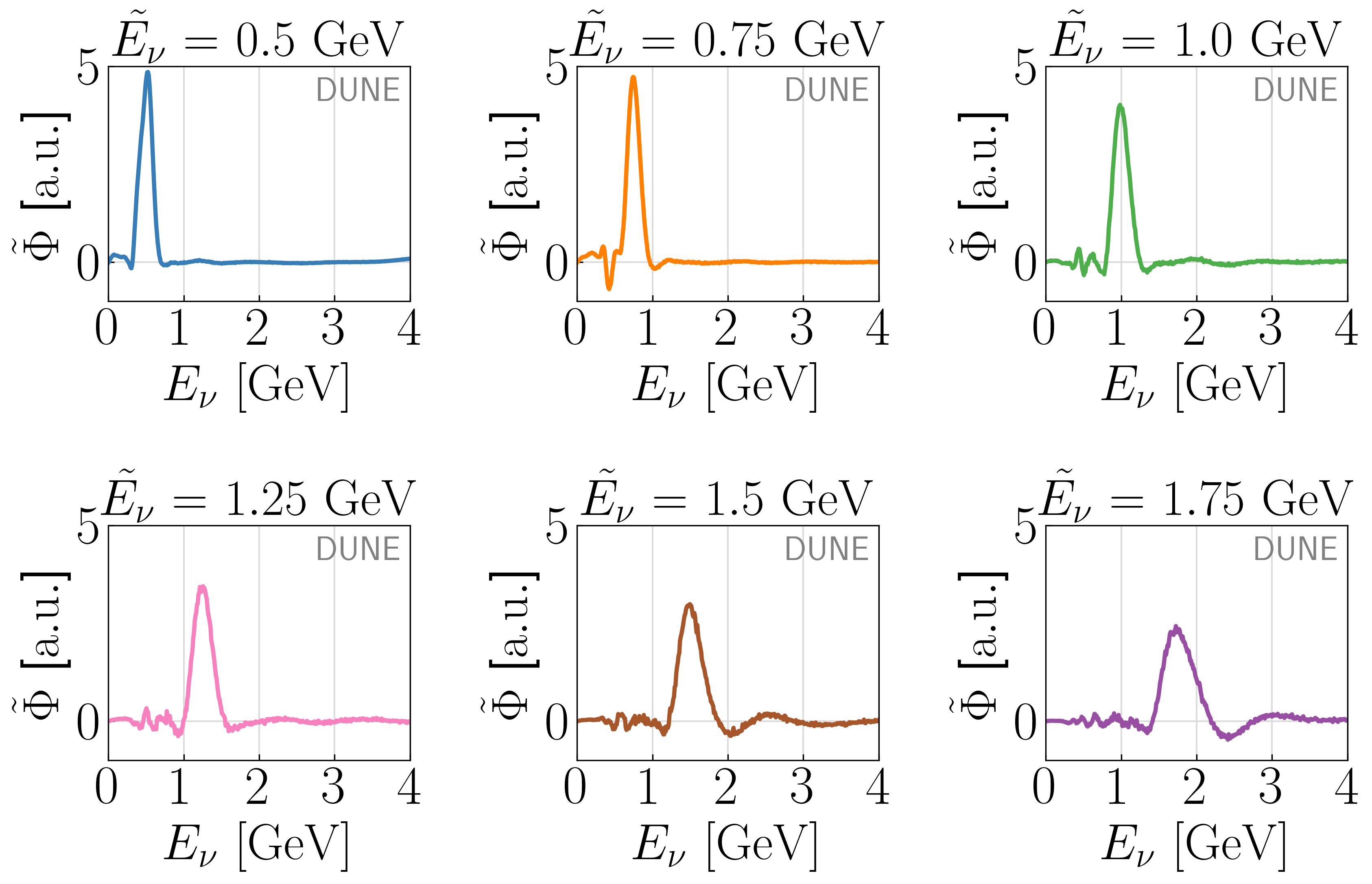}
    \caption{}
    \label{fig:IntXSecSubfig2}
  \end{subfigure}
  \caption{(a) Simulated virtual-flux-integrated charged-current muon neutrino inclusive cross-section measurements from DUNE-PRISM. The dotted line plots true \(\sigma(E_{\nu})\). 
 Dashed and hollow data points are flux integrated simulated measurements based on virtual fluxes constructed with the target fluxes according to \autoref{virtualxsec}. Colored data points are the same measurements but applying \autoref{virtualxsec_cut}, including a cut to remove high-energy contributions to the virtual flux tails. The colors correspond to the virtual fluxes shown in the right figure. The horizontal position of each point corresponds to the mean neutrino energy \(\tilde{E}_\nu\) of the virtual flux; its width in \(E_\nu\) is the standard deviation of that flux. The vertical (y-axis) uncertainty bars for each point (both colored and dashed data points) is composed of shape systematic uncertainty (inner cap) and statistical uncertainty (middle cap), as well as the total uncertainty (systematic and statistical, outer cap, lighter color), assuming $\sim$2.5 years of DUNE running (in a non-nominal run plan, see appendix \ref{app:evtrates}). The empty colored bar shows the separate normalization systematic uncertainty. The bottom panel is a residuals plot, showing the relative difference between the simulated PRISM cross-section measurements (colored points) and the true cross section (dashed line). The error bars are displayed following the same rule as the main panel, reflecting total uncertainty, including statistical and systematic components. (b) Virtual fluxes on which the simulated measurements are based, with line colors matching data point colors on the left.}
  \label{fig:IntXSec}
\end{figure*}

\begin{figure*}[tb]
    \centering
    \begin{subfigure}{0.35\linewidth}
        \centering
        \includegraphics[width=\linewidth]{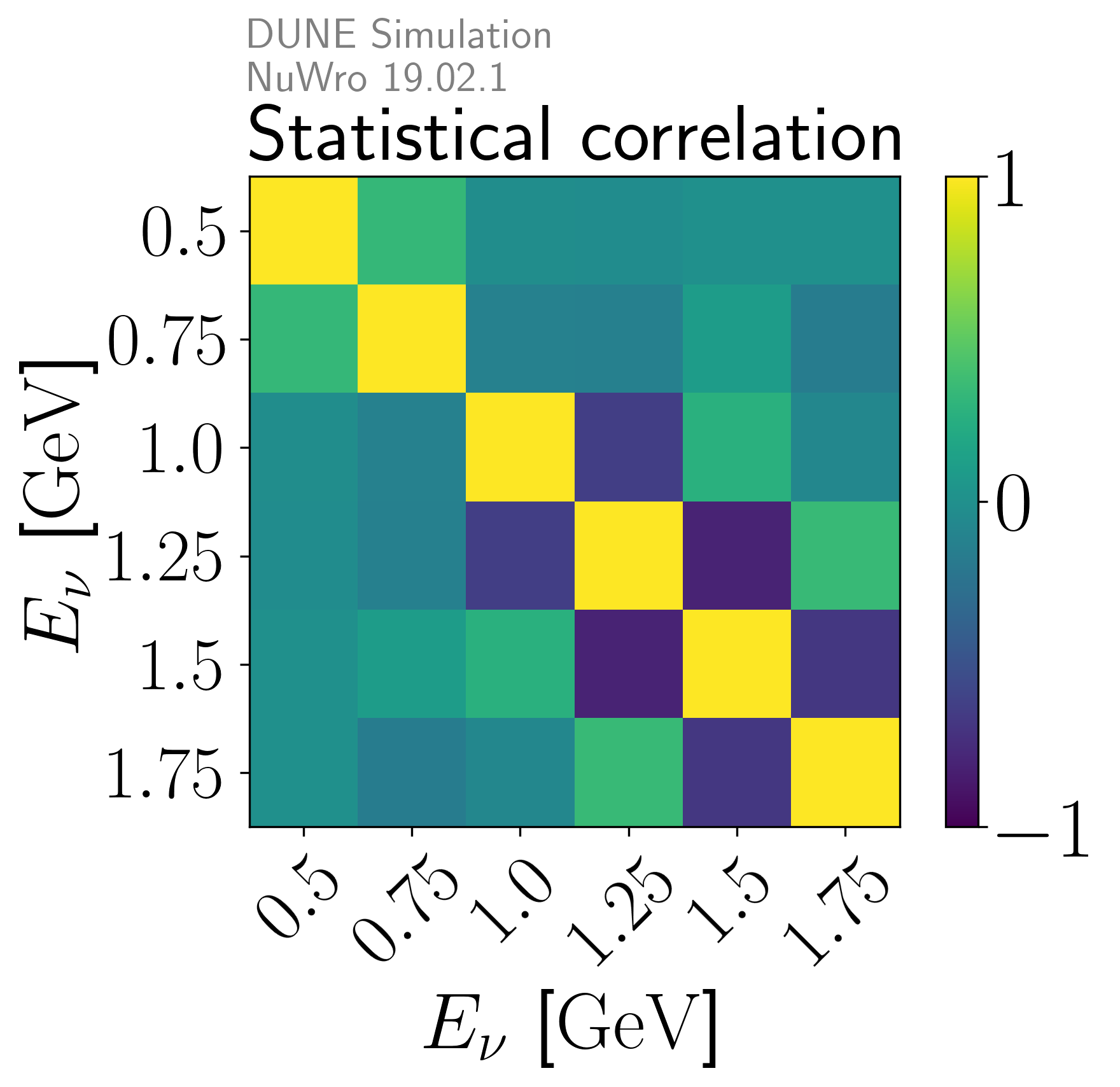}
        \caption{}
        \label{fig:IntStatCorr}
    \end{subfigure}
    \begin{subfigure}{0.35\linewidth}
        \centering
        \includegraphics[width=\linewidth]{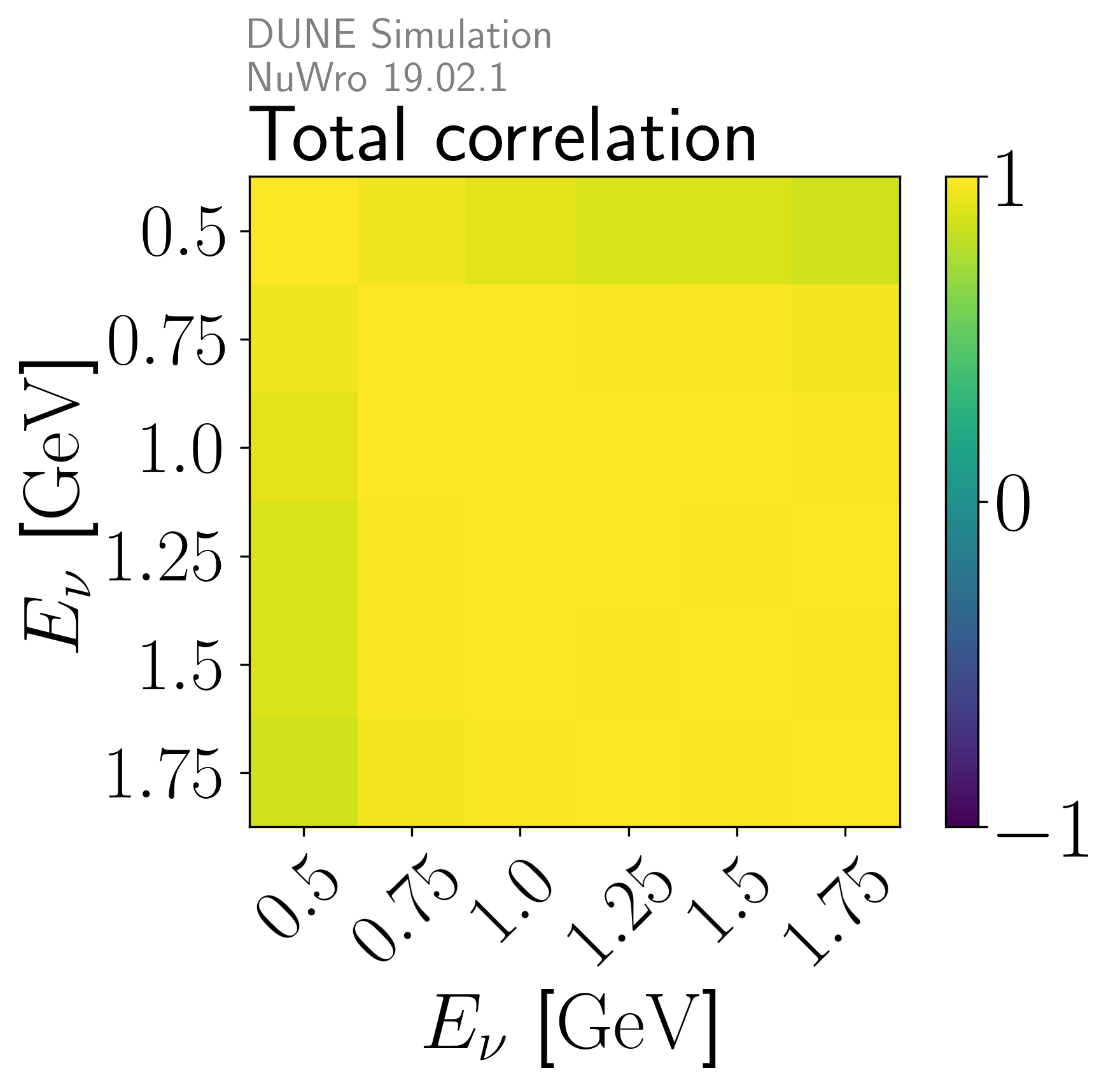}
        \caption{}
        \label{fig:IntTotalCorr}
    \end{subfigure}
    \caption{Correlations between the simulated flux-integrated cross-section measurements plotted in \autoref{fig:IntXSec}. (a) Correlation between integrated cross section measurements, calculated with toys generated based on statistical uncertainties; (b) Total uncertainty correlation, calculated by combining the covariance of the systematic toys measurements, calculated based on toys generated based on systematic uncertainties, and statistical toys measurements.}
    \label{fig:intxseccorr}
\end{figure*}


In order to explore DUNE-PRISM's potential ability to mimic monochromatic fluxes, six values of incoming neutrino energies below 2~GeV were chosen, as described in \autoref{sec:xsecwithvflux}. The desired virtual fluxes were then chosen as Gaussian distributions centered around each of these values with a standard deviation of 70~MeV. This value was chosen to balance the need for a flux that is close to monochromatic while keeping statistical uncertainties reasonable (as discussed in \autoref{sec:virtualfluxes}). However, the exact choice of target fluxes and the regularization used in the computation of the virtual flux still allow room for additional optimization.  

\autoref{fig:IntXSec} shows the six different virtual fluxes considered and the flux-averaged integrated cross section measured for each of them (with and without a cut on neutrino energy) with uncertainties assuming $\sim$2.5 years of DUNE data taking (in a non-nominal run plan, see appendix \ref{app:evtrates}). These results are compared to the true cross section, predicted by the NuWro generator, as a function of neutrino energy and the flux averaged cross section calculated using the perfect Gaussian target fluxes. Non-Gaussian features in the virtual fluxes move the projected cross sections from the ideal Gaussian flux case. As discussed in~\ref{sec:xsecwithvflux}, the most impactful of these features are present at higher energies, where small alterations to the flux are enhanced by the larger cross section. Applying the  neutrino energy cut therefore brings the virtual flux-integrated cross sections significantly closer to an ideal Gaussian shape, making the results easier to interpret and allowing virtual flux-integrated measurements to serve as reasonable estimates of the true cross section. While the cut on neutrino energy typically reduces the cross section across bins, it leads to a modest increase in the 1.25 GeV case. This is because this virtual flux has a small negative contribution at energies above 4 GeV. The remaining differences to the ideal Gaussian case stems from the features observable on the right of \autoref{fig:IntXSec} outside of the main peak.  Features to the left of the peaks are suppressed by their smaller cross section, features to the right of the peak (but below the energy cut) are in regions of larger cross section and so enhanced. However, overall these non-Gaussian features do not prevent the virtual flux-averaged cross section from being a thin a well-understood virtual flux.

While the virtual fluxes visibly deviate more from their Gaussian targets at higher energies, this is not inherently problematic. As discussed in \autoref{sec:xsecwithvflux}, the target fluxes are not physical inputs we aim to match exactly, but rather auxiliary tools used to construct virtual fluxes with desirable properties. As long as the resulting virtual fluxes remain sufficiently narrow and well-characterized, the corresponding cross-section measurements remain useful.

The observed differences between the projected flux-averaged measurements (with or without the energy cut) and those obtained using idealized Gaussian target fluxes do not indicate a flaw in the analysis. Rather, they illustrate the limitations of representing a virtual flux using only a single characteristic energy and associated uncertainty. A complete quantitative model comparison to these virtual flux integrated measurements can still be performed but the models would need to be folded through the virtual flux the measurement was made with. 

The correlations between bins in the projected measurement of the integrated cross section can be seen in \autoref{fig:intxseccorr}. The very large correlations across all bins, clearly dominated by the flux systematic uncertainties, indicate that the measurement is dominated by an overall normalization uncertainty. Considering only the uncertainty on the shape of the measurement, the combined statistical and systematic uncertainty ranges between 3.8\%-7.2\% with $\sim$2.5 years of DUNE data (in a non-nominal run plan, see appendix \ref{app:evtrates}). The total uncertainty, including normalization uncertainty ranges between 9.0-11.6\%. While utilizing the full 20-year dataset would further reduce statistical uncertainties, meaningful results with less than 10\% combined statistical and systematic shape uncertainties will be available earlier.
The statistical uncertainty induced correlations between different flux-averaged cross-section measurements are also shown in \autoref{fig:intxseccorr}, demonstrating significant correlation strength. This is expected, since all of the data points are based on the same set of events; for example, giving weights with opposite signs to the same events will increase the anti-correlation between data points.

\subsection{Differential cross-section measurements}
\begin{figure*} [tb]
    \centering
    \includegraphics[width=1\linewidth]{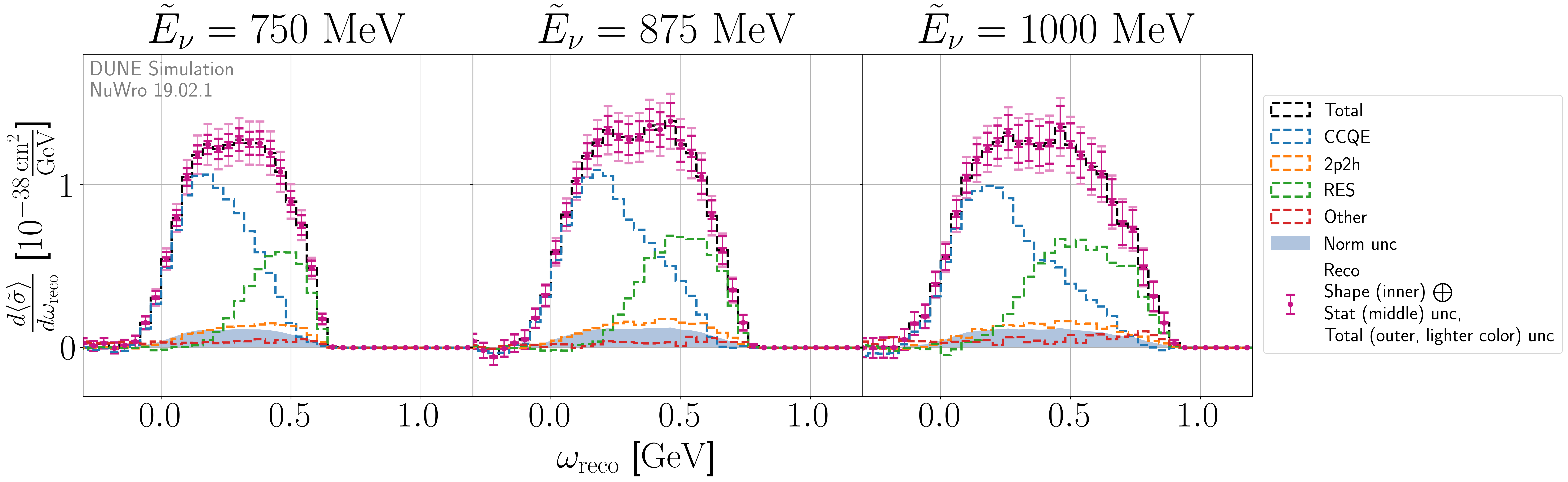}
    \caption{Simulated differential cross-section measurements of inclusive charged-current muon neutrino interactions, shown as a function of reconstructed energy transfer, for Gaussian virtual fluxes centered at different neutrino energies. The uncertainty is composed of shape systematic uncertainty (inner cap) and statistical uncertainty (middle cap), as well as the total uncertainty (systematic and statistical, outer cap, lighter color). The blue band shows the normalization systematic uncertainty separately. Colored interaction mode histograms are based on the true MC interaction classification.}
    \label{fig:recores}
\end{figure*}
\begin{figure*}[tb]
    \centering
    \begin{subfigure}{0.32\linewidth}
        \centering
        \includegraphics[width=\linewidth]{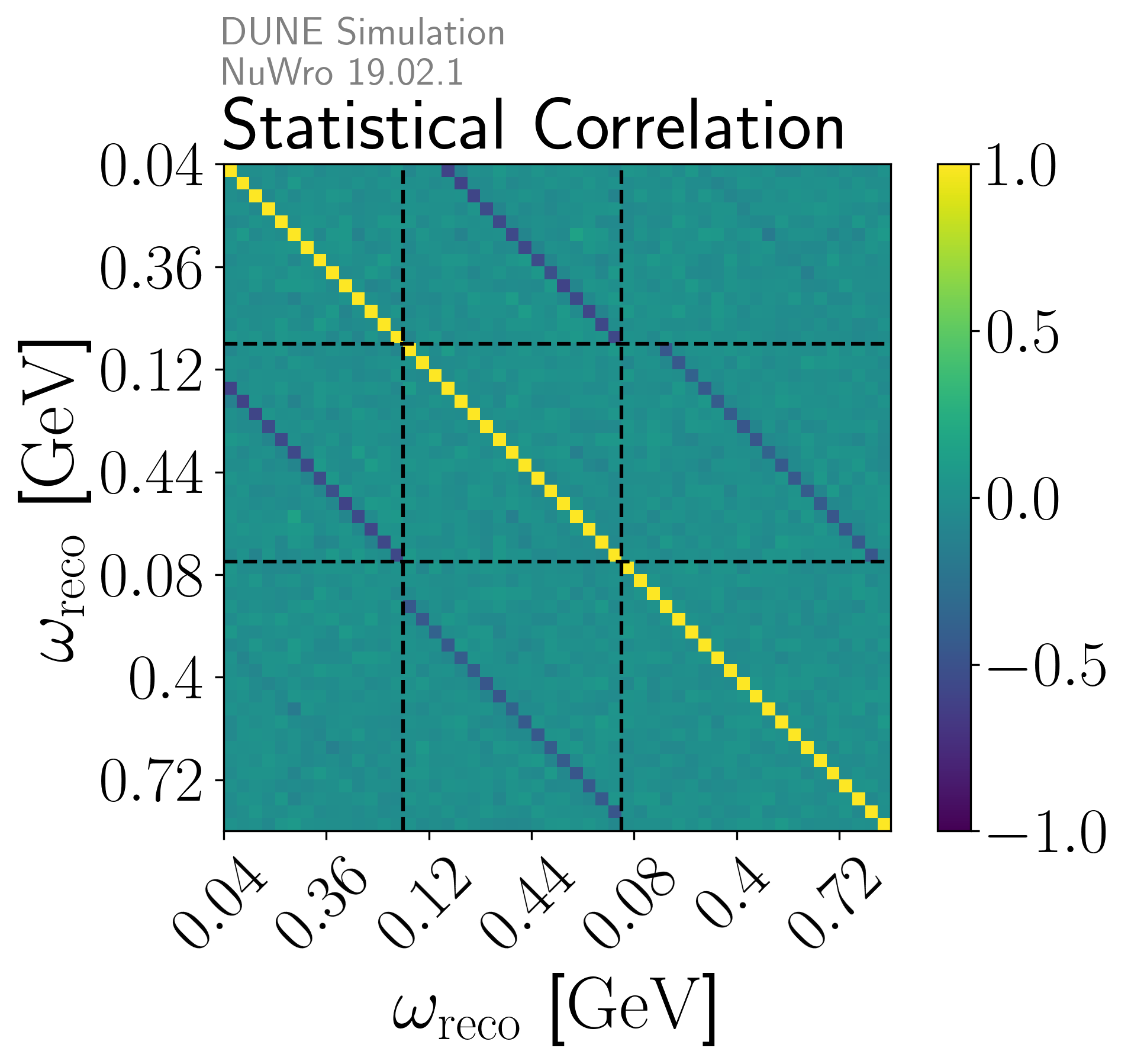}
        \caption{}
        \label{fig:DiffStatCorr}
    \end{subfigure}
    \begin{subfigure}{0.32\linewidth}
        \centering
        \includegraphics[width=\linewidth]{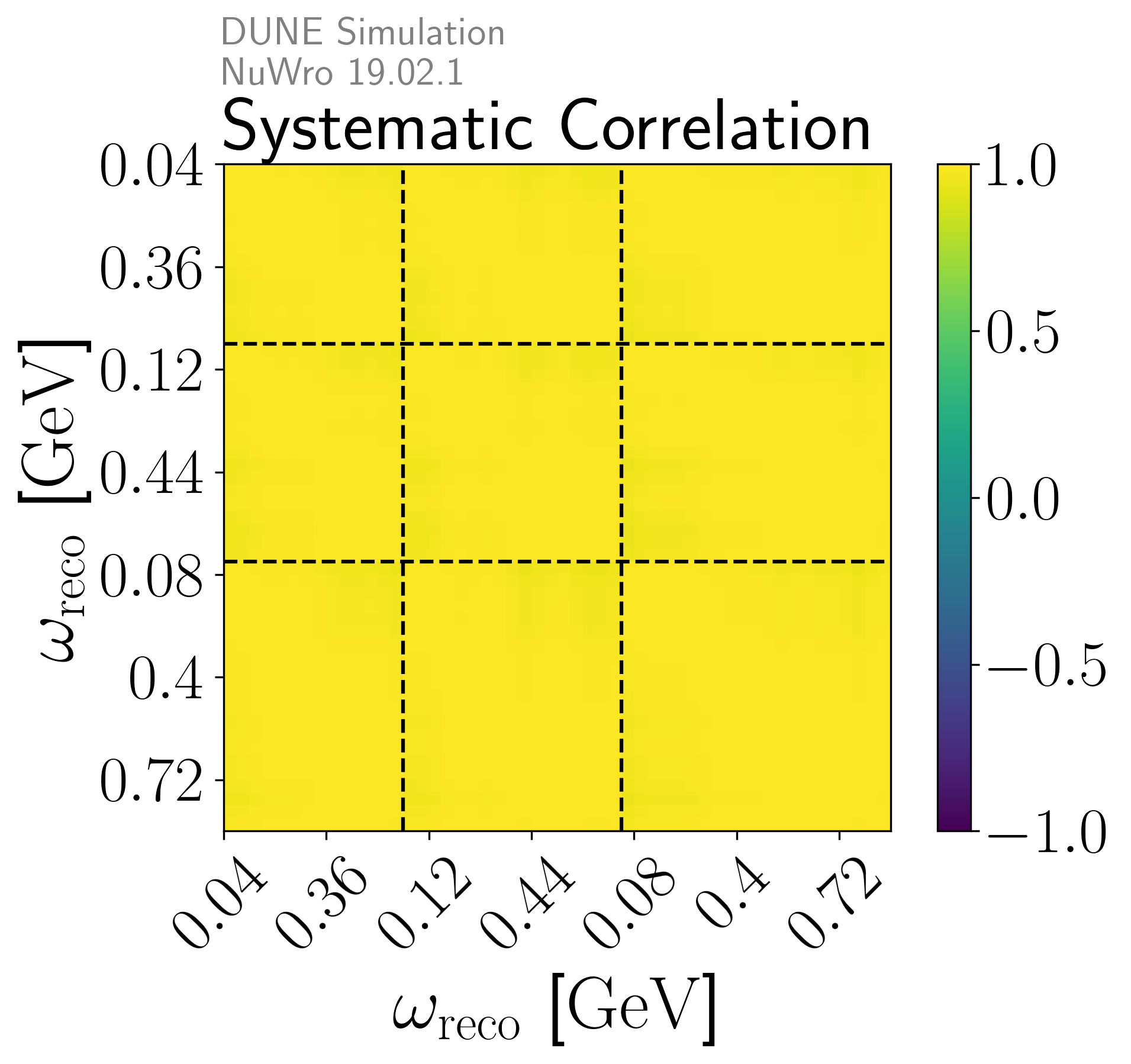}
        \caption{}
        \label{fig:DiffSysCorr}
    \end{subfigure}
    \begin{subfigure}{0.32\linewidth}
        \centering
        \includegraphics[width=\linewidth]{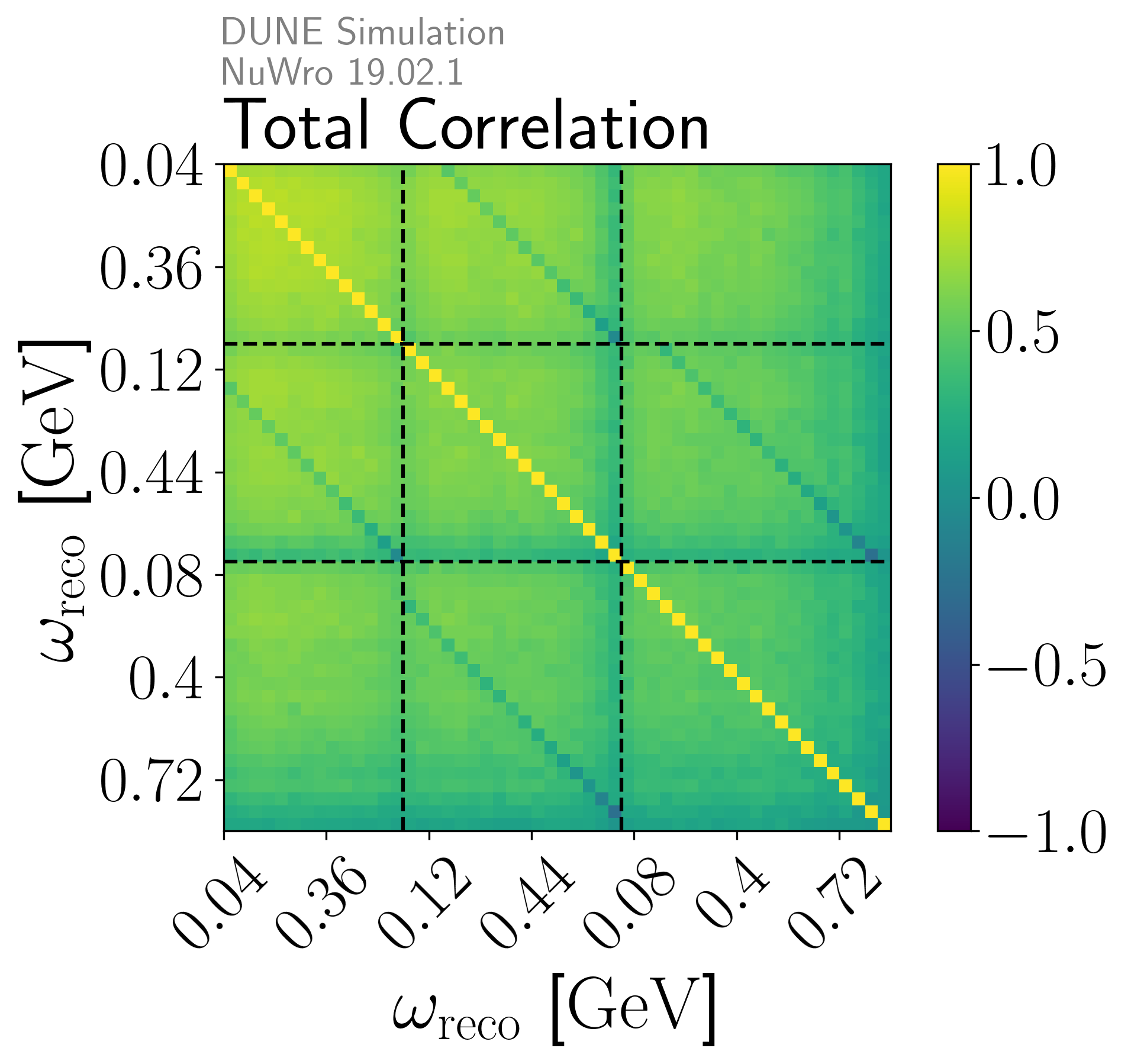}
        \caption{}
        \label{fig:DiffTotalCorr}
    \end{subfigure}
    \caption{Correlations between the simulated differential cross-section measurements plotted in \autoref{fig:recores}. (a) Statistical uncertainty correlations, assuming $\sim$10 years of DUNE running (in a non-nominal run plan, see appendix \ref{app:evtrates}); (b) Systematic uncertainty correlations; (c) Total uncertainty correlation of both systematic and statistical uncertainty. In all of the plots, the \((i,j)\) block corresponds to the correlation between the \(i\)-th and \(j\)-th leftmost simulated measurement in \autoref{fig:recores}.}
    \label{fig:recorescorr}
\end{figure*}

For three virtual fluxes, the differential cross sections are obtained as a function of reconstructed energy transfer as defined in \autoref{eq_omegareco}. 
\FigureAutoref{fig:recores} shows \(\frac{d\langle \tilde{\sigma} \rangle}{d\omega_{\mathrm{reco}}}\) compared to the true simulated distributions along with the different contributions of each charged-current reaction channels: quasi elastic (CCQE), meson exchange (2p2h), resonances (RES) and deep inelastic scattering (making up most of the ``other'' component). More details on the modeling of the different reaction channels can be found in Refs.~\cite{NuSTEC:2017hzk, Golan:2012wx}. The different CCQE and RES reaction channels make up two distinct peaks with 2p2h filling in the gap between them, as is observed in electron scattering data~\cite{Ankowski:2022thw}. 

The uncertainty of the results was estimated by assuming $\sim$10 years of DUNE data taking (in a non-nominal run plan, see appendix \ref{app:evtrates}). They are dominated by a normalization uncertainty, as seen by the blue filled area in \autoref{fig:recores}, which leads to highly correlated measurements, as seen in \autoref{fig:recorescorr}. However, excluding normalization uncertainty, the combined statistical and systematic shape uncertainty remains below 9\% in the region where the reconstructed energy transfer \(\omega_{\text{reco}}\) is kinematically allowed for a neutrino of energy \(\tilde{E}_{\nu}\), i.e., for \(\omega_{\text{reco}} \leq \tilde{E}_{\nu} - m_{\mu}\). This region corresponds to the domain in which \(\omega_{\text{reco}}\) approximates the true energy transfer \(\omega_{\text{true}}\), and the plotted values can therefore be interpreted as physically meaningful differential cross sections. 

We assess that the few highly anti-correlated bins between simulated measurements that are based on different virtual fluxes are the result of the fluxes having coefficients with opposite signs; thus, more events in one off-axis position increase the result for one flux and decrease it for the other.

\section{Discussion} \label{sec:discuss}

The results in \autoref{sec:results} demonstrate how DUNE-PRISM data, taken with a broad spectra of neutrino energies, could be utilized to estimate neutrino-nucleus cross section for specific incoming neutrino energies. 

A simulated measurement of the integrated inclusive cross section, shown in \autoref{fig:IntXSec}, resulting from six narrow virtual-flux averaged measurements, is very close to the cross section at the virtual flux mean neutrino energy, \(\tilde{E}_{\nu}\), especially when applying a cut on reconstructed neutrino energy (which can be approximated using a cut on visible neutrino energy, as described in \autoref{sec:intxsecmethods}). This implies that DUNE-PRISM is capable of measuring neutrino interaction cross sections as a function of neutrino energy. Such measurements have been attempted using broad-band fluxes~\cite{Barish:1978pj,Baker:1982ty,MicroBooNE:2021sfa,MicroBooNE:2023foc,T2K:2014hih} typically with $\sim$10\% uncertainties. At slightly higher energies but still in the DUNE neutrino energy range, $\sim$5\% precise measurements have been presented by NOMAD, MINOS, and MINERvA using the low-$\nu$ method~\cite{MINERvA:2017ozn,MINERvA:2016ing,MINOS:2009ugl}. Whilst this precision is comparable to or exceeds that shown in \autoref{fig:IntXSec}, these measurements attempt to reconstruct the neutrino energy from the kinematics of outgoing particles and so are inherently reliant on a model to relate reconstructed and true energies. Moreover, the use of the low-$\nu$ method has also been shown to exhibit considerable model dependence at GeV-scale neutrino energies~\cite{Wilkinson:2022dyx}. Conversely, a measurement of the neutrino cross section as a function of neutrino energy by DUNE-PRISM is expected to have almost no interaction model dependence. 

Concerning the differential cross sections, presented in \autoref{fig:recores}, showing simulations of measurements of \(\frac{d\langle \tilde{\sigma} \rangle}{d\omega_{\mathrm{reco}}}\), for increasing values of \(\tilde{E}_{\nu}\) the distributions become double-peaked, as RES interactions are dominant when \(\tilde{E}_{\nu}\) approaches 1~GeV. The experimentally observable distribution shows clear evidence of this underlying structure through their broadening, allowing a meaningful separation between the CCQE and RES dominant regions. This is observable even without constraining the outgoing lepton angle (as is done in electron scattering measurements). As in the case of the integrated cross section, experimental inferences of energy transfer in neutrino scattering have been made, but these are either model dependent or measure variables that are smeared with respect to true energy transfer in a complicated way (see Ref.~\cite{Wilkinson:2022dyx} for many details regarding the challenges of reconstructing energy transfer). However, for a DUNE-PRISM measurement the energy transfer is smeared only by the width of the virtual flux and the detector muon energy resolution, significantly reducing the dependence on the cross-section model used to make the measurement (at the cost of increasing the dependence on the better-controlled flux model). In appendix~\ref{app:Unfolding}, the possibility of unfolding the relatively-well-known smearing from the flux width is explored. Overall, the results show an interesting potential for DUNE-PRISM to access kinematic variables such as energy transfer in differential cross section measurements. An even better separation of interaction channels, using the reconstructed hadronic invariant mass rather than energy transfer, is demonstrated and discussed in Appendix~\ref{app:Hadronic}. 


Unlike in electron scattering analyses, the results shown in this paper are limited by the expected statistics, as they require DUNE-PRISM to run at many off-axis angles for sufficient time and the construction of virtual event rates involves subtractions of similarly-sized large numbers. We note that further optimization of the virtual flux construction algorithm described in \autoref{sec:matching} (for example, by taking systematics into account when choosing the optimal coefficients, by building fluxes with other shapes besides Gaussians or by optimizing the regularization used) as well as run plan optimization, and a wise choice of the virtual fluxes could significantly improve the results. In any case, our results suggest that integrated cross-section measurements as a function of neutrino energy could be obtained within the first few years of DUNE-PRISM running. 

It is found that flux uncertainties contribute significantly to these results. However, we observe that they primarily contribute to an overall normalization uncertainty and so are less significant in a shape-only analysis (e.g. when constraining the shape of the cross section as a function of neutrino energy or of energy transfer). 

Overall, this analysis provides access to usually inaccessible kinematic variables that characterize neutrino interactions in new ways, allowing future measurements to probe neutrino scattering at a level of precision usually only possible for electron scattering. These studies can also be expanded. In particular, it would be interesting to extend the currently simulated differential measurements, which provide results analogous to those of inclusive electron scattering experiments, to the exclusive case. In addition to updating the studies presented here to include realistic detector effects, future work could explore measurements of missing momentum and energy, expanding on Refs.~\cite{JeffersonLabHallA:2022cit,JeffersonLabHallA:2022ljj,MicroBooNE:2023krv}, as well as exploring the possibility of measurements of the difference between true and reconstructed neutrino energy, which is a key ingrediant for DUNE's neutrino oscillation physics program. 

\section{Conclusions}
\label{sec:concl}

This work represents a first feasibility study towards demonstrating that DUNE-PRISM can be used to make cross-section measurements averaged over virtual fluxes that are narrower than key features of cross section, such as the CCQE and RES peaks in energy transfer or the hadronic invariant mass. An analysis of simulated measurements is shown to offer unique and largely model-independent access to the cross section as a function of neutrino energy as well as differential cross sections as a function of these observables, usually reserved for electron scattering experiments. Whilst the current study suggests statistical uncertainties will be significant, there remains considerable scope for optimization in the analysis, particularly with regards to the choice of run plan and of the specific target fluxes and regularisation used to build the virtual fluxes.

\section*{Acknowledgements}

This document was prepared by DUNE collaboration using the resources of the Fermi National Accelerator Laboratory (Fermilab), a U.S. Department of Energy, Office of Science, Office of High Energy Physics HEP User Facility. Fermilab is managed by Fermi Forward Discovery Group, LLC, acting under Contract No. 89243024CSC000002.
%
%
This work was supported by
CNPq,
FAPERJ,
FAPEG and 
FAPESP,                         Brazil;
CFI, 
IPP and 
NSERC,                          Canada;
CERN;
M\v{S}MT,                       Czech Republic;
ERDF, FSE+,
Horizon Europe, 
MSCA and NextGenerationEU,      European Union;
CNRS/IN2P3 and
CEA,                            France;
INFN,                           Italy;
FCT,                            Portugal;
NRF,                            South Korea;
Generalitat Valenciana, 
Junta de AndalucÄ±a-FEDER, 
MICINN, and 
Xunta de Galicia,               Spain;
SERI and 
SNSF,                           Switzerland;
T\"UB\.ITAK,                    Turkey;
The Royal Society and 
UKRI/STFC,                      United Kingdom;
DOE and 
NSF,                            United States of America.

This work is supported by ERC grant (NeutrinoNuclei, 101078772)

\appendix

\section{Event rates}
\label{app:evtrates}

The simulation for the analysis within this manuscript assumes equal exposure in each off-axis position from on-axis to 3.32 degrees off axis, in slices of 1 mrad (approximately 0.057 degrees; a 0.574 m wide slice). For comparison, a preliminary DUNE yearly run plan in Ref.~\cite{DUNE:2021tad} assumes 12 days for 7 different off-axis positions, alongside a 14 week exposure on-axis plus another week on-axis with an alternative horn current. Therefore, our exposure in the nominal plan would correspond to 90 out of the 201 days of data taking per year (the 7 times 12 day DUNE-PRISM off-axis positions plus 6 days of an on-axis position, which samples the off-axis angles on either side of the beam). 

The event rates in each off-axis slice considered in the manuscript per 90 days are shown in \autoref{tab:evtrates}. One year of DUNE exposure in the DUNE preliminary run plan would provide these event rates (alongside many more events on-axis). In a dedicated non-nominal plan, where each position was sampled evenly, this would correspond to a little less than half of a year of data taking. Throughout this manuscript we give exposure times based on these event rates corresponding to half a year of data. 

\begin{table*}[]
\begin{tabular}{l|l|l|l}
Off-axis index & Lower edge {[}deg.{]} & Upper edge {[}deg.{]} & Events / 90 days \\ \hline
1& 0.00& 0.06& 299584.76     \\
2& 0.06& 0.11& 298570.86     \\
3& 0.11& 0.17& 295320.91     \\
4& 0.17& 0.23& 290926.46     \\
5& 0.23& 0.29& 283970.18     \\
6& 0.29& 0.34& 275706.26     \\
7& 0.34& 0.40& 264759.94     \\
8& 0.40& 0.46& 251676.87     \\
9& 0.46& 0.52& 237966.82     \\
10& 0.52& 0.57& 222610.65     \\
11& 0.57& 0.63& 206338.86     \\
12& 0.63& 0.69& 189282.06     \\
13& 0.69& 0.74& 172244.09     \\
14& 0.74& 0.80& 155980.15     \\
15& 0.80& 0.86& 140945.11     \\
16& 0.86& 0.92& 127201.72     \\
17& 0.92& 0.97& 114669.57     \\
18& 0.97& 1.03& 103686.96     \\
19& 1.03& 1.09& 93687.16      \\
20& 1.09& 1.15& 85014.86      \\
21& 1.15& 1.20& 77544.09      \\
22& 1.20& 1.26& 70862.19      \\
23& 1.26& 1.32& 64983.13      \\
24& 1.32& 1.38& 59823.02      \\
25& 1.38& 1.43& 55173.42      \\
26& 1.43& 1.49& 51090.73      \\
27& 1.49& 1.55& 47412.73      \\
28& 1.55& 1.60& 44035.92      \\
29& 1.60& 1.66& 41045.33      \\
30& 1.66& 1.72& 38331.74      \\
31& 1.72& 1.78& 35822.96      \\
32& 1.78& 1.83& 33523.61      \\
33& 1.83& 1.89& 31451.68      \\
34& 1.89& 1.95& 29516.61      \\
35& 1.95& 2.01& 27786.59      \\
36& 2.01& 2.06& 26156.78      \\
37& 2.06& 2.12& 24639.36      \\
38& 2.12& 2.18& 23189.62      \\
39& 2.18& 2.23& 21958.92      \\
40& 2.23& 2.29& 20703.17      \\
41& 2.29& 2.35& 19536.93      \\
42& 2.35& 2.41& 18526.06      \\
43& 2.41& 2.46& 17537.77      \\
44& 2.46& 2.52& 16565.03      \\
45& 2.52& 2.58& 15708.49      \\
46& 2.58& 2.64& 14906.22      \\
47& 2.64& 2.69& 14130.65      \\
48& 2.69& 2.75& 13418.87      \\
49& 2.75& 2.81& 12784.37      \\
50& 2.81& 2.86& 12136.05      \\
51& 2.86& 2.92& 11540.08      \\
52& 2.92& 2.98& 10997.79      \\
53& 2.98& 3.04& 10482.17      \\
54& 3.04& 3.09& 10007.06      \\
55& 3.09& 3.15& 9544.40       \\
56& 3.15& 3.21& 9107.26       \\
57& 3.21& 3.27& 8691.73       \\
58& 3.27& 3.32& 8313.94       \\ \hline
Total          & 0.00                  & 3.32& 5.17 M    \\
\end{tabular}
\caption{The angular range and event rates considered for each of the 58 off-axis slices considered in our analyses. The event rates correspond to a total of 90 days of DUNE-PRISM data taking (just under half a DUNE beam year) in a non-nominal run plan where each off-axis position is sampled evenly. The same events would be available as a subset of data taken in one year of DUNE running under the preliminary run plan given in Ref.~\cite{DUNE:2021tad}.}
\label{tab:evtrates}
\end{table*}

\section{Unfolding to true energy transfer}
\label{app:Unfolding}

\begin{figure} [tb]
    \centering
    \includegraphics[width=\linewidth]{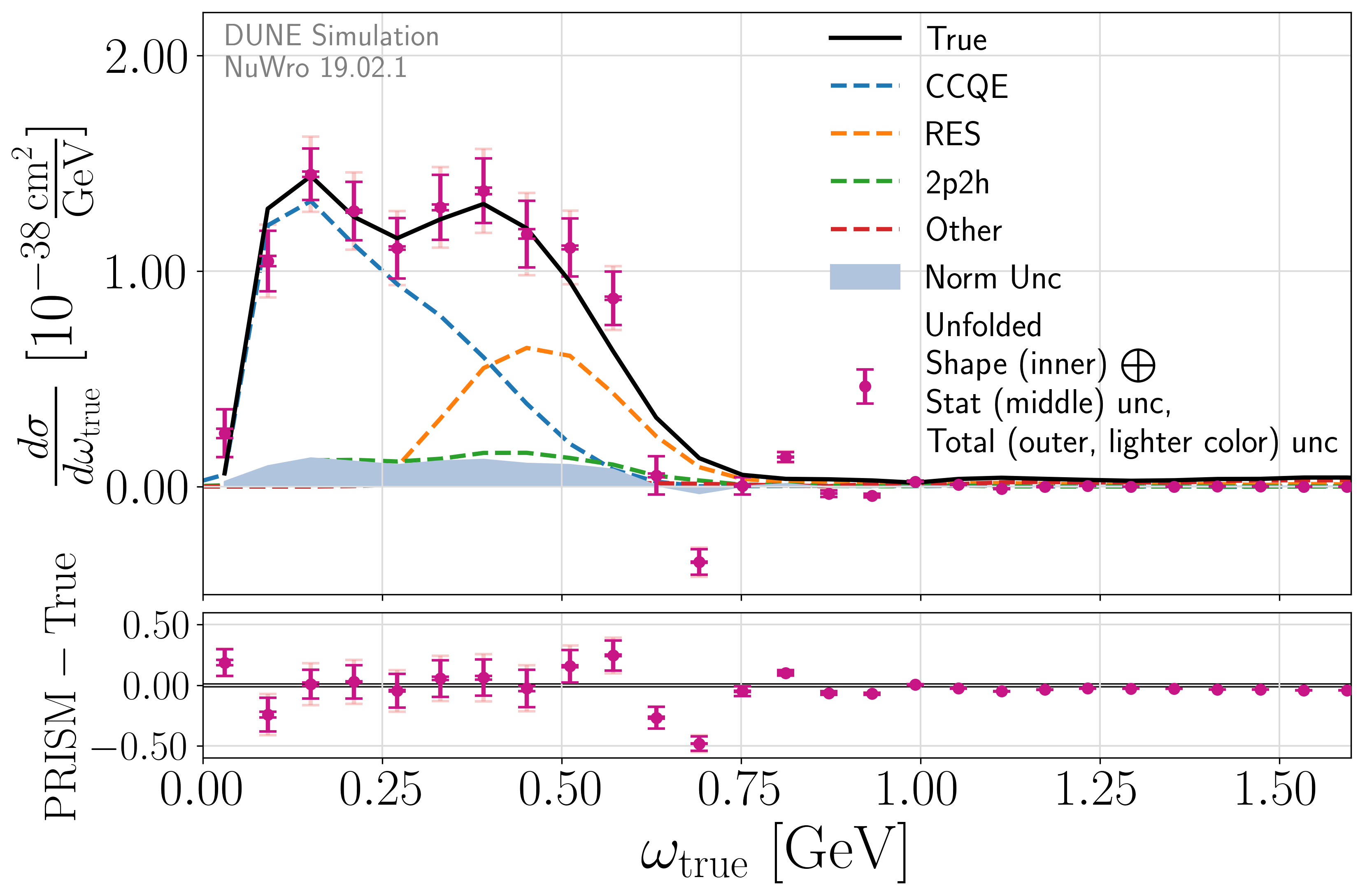}
    \caption{Unfolded simulated differential cross-section measurement as a function of the true energy transfer, \(\omega_{\text{true}}=E_{\nu}-E_{\mu}\), for a Gaussian target flux with \(\tilde{E}_{\nu}=0.75\) GeV and  flux width of \(78\) MeV (with a \(70\) MeV target width). Violet data points are the unfolded differential cross section simulated measurement, calculated by unfolding \(\frac{d\langle\sigma\rangle}{d\omega_{\text{reco}}}\). The uncertainty is composed of statistical uncertainty (inner cap) and shape systematic uncertainty (middle cap), as well as total uncertainty (systematic and statistical, outer cap, lighter color). The blue band shows the separate normalization systematic uncertainty. Statistical uncertainties assume $\sim$10 years of DUNE-PRISM data taking (in a non-nominal run plan, see appendix \ref{app:evtrates}). The black line is the interpolated true differential cross section. Colored interaction mode histograms are based on the true simulation interaction classification. The bottom panel is a residuals plot, showing
    the absolute difference between the simulated unfolded PRISM cross-section measurements (violet points) and the true cross differential cross section (vertical line).}
    \label{fig:UnfoldedExample}
\end{figure}

Virtual flux based measurements of differential cross sections as a function \(\omega_{\mathrm{reco}}\), can in principle be unfolded into differential cross sections as a function of the true energy transfer, \(\omega_{\mathrm{true}}=E_{\nu}-E_{\mathrm{lepton}}\). This can be done in a minimally model-dependent way since the smearing is dominated by the well-known width of the virtual flux used to make the measurements. Since we are only considering flux uncertainties, \(\omega_{\mathrm{reco}}\) only differs from \(\omega_{\mathrm{true}}\) according to the energy spread of the virtual flux, and therefore we model \(S\) using a Gaussian smearing matrix constructed from the spread in the virtual flux.

Event rates in both variables are related by:

\[\left(\frac{d\tilde{N}}{d\omega_{\mathrm{reco}}}\right)_{i}=\sum_{i,j} S_{ij} \left( \frac{d\tilde{N}}{d\omega_{\mathrm{true}}}\right)_{j},\]

\noindent where the smearing matrix component \(S_{ij}\) is the probability of an event in \(\omega_{\mathrm{true}}\) bin \(j\) to be reconstructed in \(\omega_{\mathrm{reco}}\) bin \(i\). For a narrow enough flux \(\tilde{\Phi}(E_{\nu})\) centered at \(\tilde{E}_{\nu}\), where \(\tilde{\Phi}(E_{\nu})\approx \delta(E_{\nu}-\tilde{E}_{\nu})\), \(\sigma(E_{\nu}\)) is approximately constant around \(\tilde{E}_{\nu}\). Thus, \(\frac{d\tilde{N}}{dE_{\nu}}\propto\tilde{\Phi}(E_{\nu})\), which allows us to assume a Gaussian smearing matrix or one that is directly constructed using \(\tilde{\Phi}(E_{\nu})\). We explored both a Gaussian smearing matrix and one constructed directly from \(\tilde{\Phi}(E_{\nu})\), and found that the former performed better in our study.

To unfold the simulated measurements, we used a least squares algorithm with Tikhonov regularization. The regularization parameter for each distribution was selected using an L-curve method. The algorithm we used finds the point of maximal curvature with a trade-off between the unfolded distribution's \(L_2\) norm, which regularizes the distribution's variance (to mitigate oscillations), and its integrated value, which regularizes its total bias (to avoid results that are too suppressed). This algorithm was applied separately to each toy \(\frac{d\tilde{N}}{d\omega_{\mathrm{reco}}}\) distribution. Uncertainties were evaluated by calculating the covariance of these toy unfolded distributions, for the statistical and systematic uncertainties separately (in the same way they were calculated for \(\frac{dN}{d\omega_{\mathrm{reco}}}\)).

For the Gaussian smearing matrices, the Gaussian distribution was defined such that its width is equal to the standard deviation of the virtual flux, rather than the standard deviation of the perfect target flux, mainly because virtual fluxes that are generated using Tikhonov regularization are broader than the initial target flux. The resulting cross section, using a Gaussian smearing matrix based on the flux width, is shown in \autoref{fig:UnfoldedExample}. It captures the dominant features of the true distribution, including visible CCQE and resonance (RES) peaks, demonstrating that the essential structure of the interaction dynamics is retained. At the same time, some points exhibit significant deviations from the true curve, most notably at \(\omega \sim 0.7\)~GeV and \(\omega \sim 0.9\)~GeV . These discrepancies reflect bias introduced by the unfolding procedure; in particular, regularization suppresses statistical fluctuations to stabilize the integrated result, but this comes at the cost of local accuracy. A more refined treatment of this tradeoff is left to future work.

In this unfolding procedure, the smearing matrix is built from the flux width rather than the actual smearing between true and reconstructed energy transfer. These are only exactly equivalent when the cross section is constant across the width of the flux, which is only approximately true (the narrower the virtual flux, the better this holds). This results in an extracted cross section that is slightly biased with respect to the truth. A correction for this bias would be model dependent, relying on the knowledge of the evolution of the cross section over the width of the flux, but it is also small. Methods to further evaluate and correct for this bias are left to future work.


\section{Measurements as a function of hadronic invariant mass}
\label{app:Hadronic}

\begin{figure} [tb]
    \centering
    \includegraphics[width=\linewidth]{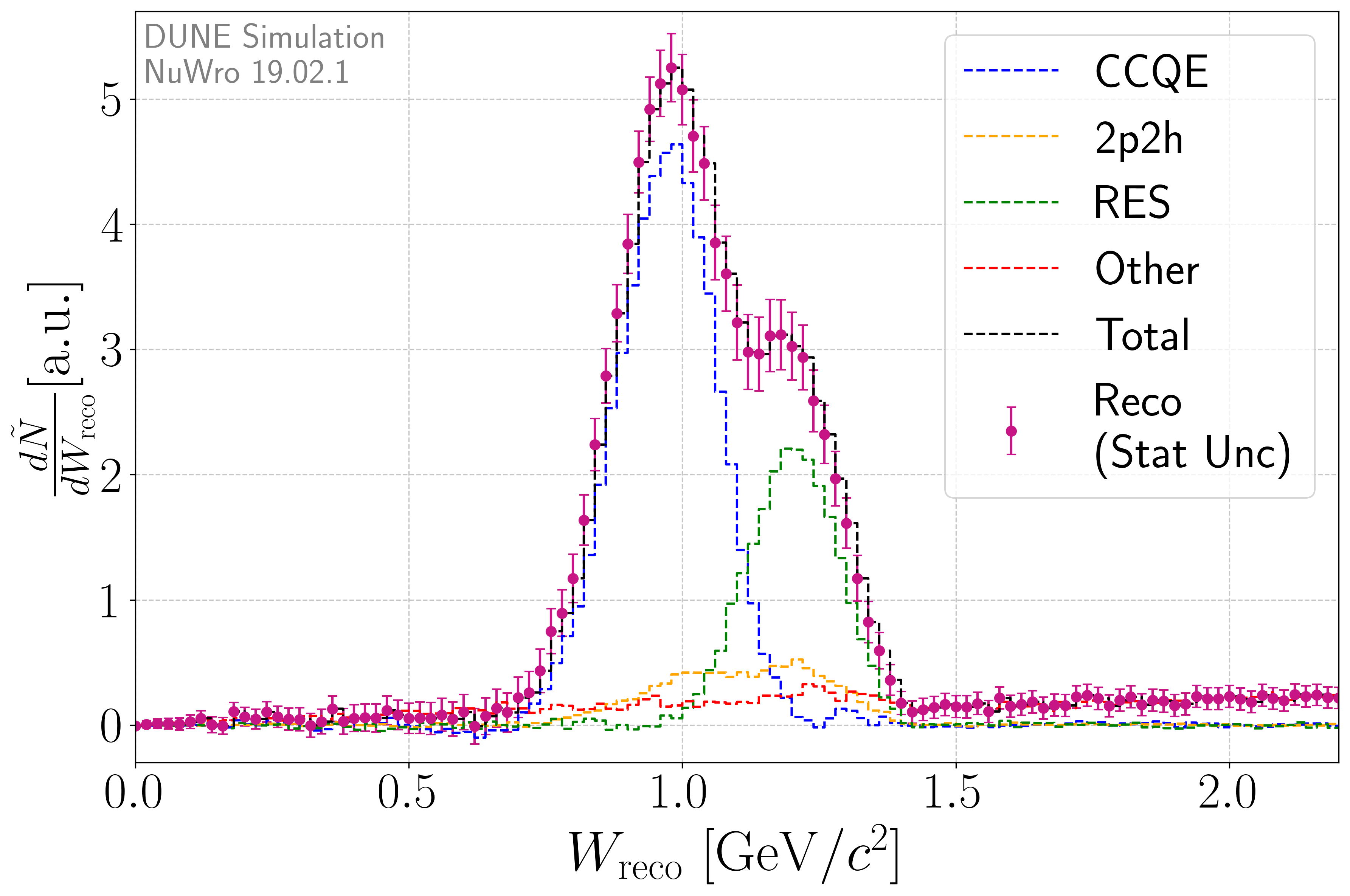}
    \caption{Simulated virtual event rates as a function of hadronic mass for a Gaussian target flux with $ \tilde{E}_{\nu}=0.75$ GeV and flux width of 78~MeV (with a 70~MeV target width). Black data points are the total projected events rate showing also the statistical uncertainty. Colored interaction mode histograms are based on the true simulated interaction classification. The statistical uncertainty is calculated assuming $\sim$10 years of data (in a non-nominal run plan, see appendix \ref{app:evtrates}).}
    \label{fig:wxsec}
\end{figure}

Analogously to the case of energy transfer detailed in \autoref{sec:xsecwithvflux}, we can additionally extract the differential cross section as a function of the reconstructed hadronic invariant mass \(\frac{d\sigma}{dW_{\mathrm{reco}}}\) where:

\begin{equation}
\label{eq_Wreco}
    W_{\mathrm{reco}}= \sqrt{M_N^2 + 2M_N(\tilde{E}_{\nu} - E_{lepton}) - Q_{\mathrm{reco}}^2},
\end{equation}

\noindent where $M_N$ is a nucleon mass, $\tilde{E}_{\nu}$ (as in \autoref{eq_omegareco}) is a proxy for true neutrino energy taken to be the mean of the target flux considered, $E_{lepton}$ is the energy of the outgoing lepton and $Q_{\mathrm{reco}}^2$ is a reconstructed version of the four-momentum transfer assuming $(E_{\nu}\approx \tilde{E}_{\nu})$:

\begin{equation}
\label{eq_Q2reco}
   Q_{\mathrm{reco}}^2= -M_{lepton}^2 + 2 \tilde{E}_{\nu} (E_{lepton} - p_{lepton} \cos{\theta_{lepton}}) ,
\end{equation}

\noindent where $p_{lepton}$, $\theta_{lepton}$ and $M_{lepton}$ are the momentum, angle with respect to the incoming neutrino and the mass of the outgoing lepton (in this case a muon) respectively. As in the case for energy transfer, for a narrow virtual flux, \(E_{\nu}\approx \tilde{E}_{\nu}\), the reconstructed hadronic invariant mass approaches its true value. The interest of this variable lies in its ability to isolate the rest mass of the baryon excited by the interaction, i.e a nucleon for CCQE interactions or a resonance (dominantly the $\Delta$ baryon with a mass of 1232 MeV) for RES interactions. 

\FigureAutoref{fig:wxsec} shows the virtual event rate as a function of $W_{\mathrm{reco}}$ compared to the true simulated distributions, along with the different contributions of each charged-current reaction channel (analogously to \autoref{fig:recores}). In this appendix, only statistical uncertainties are considered, although it is expected that the size of the flux uncertainty would be similar to the case of energy transfer. The figure demonstrates that a DUNE-PRISM measurement of $W_{\mathrm{reco}}$ clearly isolates the different interaction channels, showing a separation of the CCQE and RES peaks with 2p2h filling in the dip region between them. The peaks are consistent with their expected values (the nucleon and $\Delta$ masses). Such a measurement of $W_{\mathrm{reco}}$ is complementary to that of $\omega_{\mathrm{reco}}$, better isolating the different interaction channels but offering sensitivity to different physics.  

\section{Virtual flux equation}


\label{app:Virtual flux equation}
In this appendix, we prove \autoref{virtualxsec}, noting that the virtual flux-averaged cross section should be built from a virtual event rate rather than from a linear combination of cross-section measurements.

By definition, the flux-averaged cross section, \(\sigma\), is given as:

\begin{equation} \label{fluxaveragedxsec}
\braket{\sigma} = \frac {\int \Phi(E_{\nu})\sigma (E_{\nu})dE_{\nu}}{\int \Phi(E_{\nu})dE_{\nu}},
\end{equation}

\noindent The flux-averaged cross section for a single off-axis position, \(\sigma_{\alpha}\), is:

\begin{equation} \label{fluxaveragedxsecbins}
\braket{\sigma}_{\alpha} = \frac {\int \Phi_{\alpha}(E_{\nu})\sigma (E_{\nu})dE_{\nu}}{\int \Phi_{\alpha}(E_{\nu})dE_{\nu}},
\end{equation}

\noindent To form the virtual cross section, $\braket{\tilde{\sigma}}$, we can insert the virtual flux from \autoref{virtualflux}. If we change the order of integration and summation:

\begin{equation} \label{virtual_flux_avg}
\braket{\tilde{\sigma}} = \frac {\int \tilde{\Phi}(E_{\nu})\sigma (E_{\nu})dE_{\nu}}{\int \tilde{\Phi}(E_{\nu})dE_{\nu}} = \frac {\sum_{\alpha} \int c_{\alpha} \Phi_{\alpha}(E_{\nu})\sigma (E_{\nu})dE_{\nu} } {\int \tilde{\Phi}(E_{\nu})dE_{\nu}},
\end{equation}

\noindent If we multiply each addend in the numerator by \(\frac{\int \Phi_{\alpha}(E_{\nu})dE_{\nu}}{\int \Phi_{\alpha}(E_{\nu})dE_{\nu}}\), we get:

\begin{equation} \label{multiplied_numerator}
\braket{\tilde{\sigma}} = \sum_{\alpha} c_{\alpha} \frac{\int\Phi_{\alpha}(E_{\nu})\sigma(E_{\nu})dE_{\nu}}{\int \Phi_{\alpha}(E_{\nu})dE_{\nu}} \frac{\int \Phi_{\alpha}(E_{\nu})dE_{\nu}}{\int \tilde{\Phi}(E_{\nu})dE_{\nu}},
\end{equation}

\noindent The first fraction in each addend is our definition of \(\braket{\sigma}_{\alpha}\) in \autoref{fluxaveragedxsecbins}, so that:

\begin{equation} \label{summation_virtual}
\braket{\tilde{\sigma}}=\frac{1}{\int \tilde{\Phi}(E_{\nu})dE_{\nu}}\sum_{\alpha} c_{\alpha} \braket{\sigma}_{\alpha} \int \Phi_{\alpha}(E_{\nu})dE_{\nu},
\end{equation}

\noindent and because the following equation holds for each off-axis position:

\begin{equation} \label{event_number}
N =\mathcal{E}\cdot\varepsilon\cdot N_{\rm targets}\cdot\braket{\sigma}\cdot\int\Phi(E_{\nu})dE_{\nu},
\end{equation}

\noindent we can substitute \(\frac{N_{\alpha}}{\mathcal{E}\cdot\varepsilon\cdot N_{\rm targets}} =\braket{\sigma}_{\alpha} \int \Phi_{\alpha}(E_{\nu})dE_{\nu}\) to get:

\begin{equation} \label{virtual_cross_sec}
\braket{\tilde{\sigma}}=\frac{1}{\int \tilde{\Phi}(E_{\nu})dE_{\nu}}\sum_{\alpha} c_{\alpha} \frac{N_{\alpha}}{\mathcal{E}\cdot\varepsilon\cdot N_{\rm targets}},
\end{equation}

\noindent And finally, because \(\mathcal{E}\) and \(N_{\rm targets}\) should be separately constant for different off-axis positions, and by inserting our definition of \(\tilde{N}\) in \autoref{virtualeventrates}, we arrive at:

\begin{equation} \label{final_virtual_sigma}
\braket{\tilde{\sigma}}=\frac{\tilde{N}}{\mathcal{E}\cdot\varepsilon\cdot N_{\rm targets}\cdot \int\tilde\Phi(E_{\nu})dE_{\nu}},
\end{equation}

\noindent which is exactly \autoref{virtualxsec}. 

If we plug \autoref{virtual_flux_avg} for the case of a virtual flux in \autoref{virtualxsec}, we get:

\begin{equation} \label{virtual_event_rate}
\tilde{N} = \mathcal{E}\cdot\varepsilon\cdot N_{\rm targets} \cdot \int \tilde\Phi(E_{\nu})\sigma(E_{\nu})dE_{\nu},
\end{equation}

\noindent which is the standard relation between event rates and flux. From this equation, we can deduce that in terms of event rates, a linear combination of fluxes is completely equivalent to a real physical flux with the same distribution. It follows that by constructing a virtual flux such that $\tilde{\Phi}(E_{\nu}) \to \delta(E_{\nu}-E_{\nu}')$, we get:

\begin{equation} \label{delta_function_flux}
\sigma(E_{\nu}') \approx\frac{\tilde{N}}{\mathcal{E}\cdot\varepsilon\cdot N_{\rm targets}},
\end{equation}

\noindent so that with a narrow enough virtual flux, we can measure neutrino-nucleus interactions' cross section at a specific energy \(E_{\nu}'\). 

\bibliography{references}

\end{document}